
\input harvmac.tex

\def\pt{\partial}
\def\lt{\left}
\def\rt{\right}
\def\perpp{{\scriptscriptstyle\perp}}
\def\kb{k_{\scriptscriptstyle\rm B}}
\def\nablab{{\bf\nabla}}
\def\tiny{\scriptscriptstyle}
\def\pb{{\bf p}}

\def\n{{n}}

\def\P{{\bf P}}
\def\V{{V}}
\def\dnb{\delta{\bf n}}
\def\rb{{\bf r}}
\noblackbox
\hyphenation{log-a-rithmic calc-u-lation}
\def\half{{1\over 2}}

\def\ham{\hbox{$\cal H$}}
\def\lag{\hbox{$\cal L$}}
\def\bo#1{\hbox{$O({#1})$}}
\def\lb{\hbox{$\bar\lambda_1$}}

\def\lan{\hbox{$\bar\lambda_2$}}

\def\vb{\hbox{$\bar v$}}
\def\ofl{(\ell )}

\def\prop#1#2{{#1^i#1^j/#1^2\over K_1#1^2
+ K_3#2^2} + {\delta^{ij} - #1^i#1^j/#1^2\over K_2#1^2+K_3#2^2}}
\def\dn{\hbox{$\delta n$}}

\def\vev#1{\hbox{$\langle\,{#1}\,\rangle$}}
\def\psif{\hbox{$\psi (r,t)$}}

\def\blt{\hbox{$\bullet$}}
\def\gap{\vskip 2pt}
\def\point{\gap\item{\blt}}
\def\der#1{\hbox{$\displaystyle{d{#1}(\ell )\over d\ell}$}}
\def\derr#1{\hbox{$\displaystyle{d{#1}\over d\ell}$}}
\def\bint{\int\!\!}

\Title{}{Theory of Directed Polymers}

\centerline{Randall D. Kamien, Pierre Le Doussal\footnote{$^\dagger$}
{On leave from Laboratoire de Physique
Theorique de l'Ecole Normale Superieure Paris.} and David R. Nelson}
\bigskip\centerline{Lyman Laboratory of Physics}
\centerline{Harvard University}\centerline{Cambridge, MA 02138}

\vskip .3in
We develop a theory of polymers in a nematic solvent
by exploiting an analogy with two-dimensional quantum bosons at zero
temperature.  We argue that the theory should also describe
polymers in an {\sl isotropic} solvent.
The dense phase is analyzed in a Bogoliubov-like
approximation, which assumes a broken symmetry in the phase of the
boson order parameter.  We find
a stiffening of the longitudinal fluctuations of the nematic field,
calculate the density-density correlation function, and
extend the analysis to
the case of ferro- and electrorheological fluids.  The boson
formalism is used to derive a simple hydrodynamic theory which
is indistinguishable from the corresponding theory of polymer nematics in
an isotropic solvent at long wavelengths.  We also use
hydrodynamics to discuss the physical meaning of the boson order parameter.
A renormalization group treatment in the dilute limit shows that
logarithmic
corrections to polymer wandering, predicted by de Gennes, are unaffected
by interpolymer interactions.  A continuously variable Flory exponent
appears for polymers embedded in a {\sl two}-dimensional nematic solvent.
We include free polymer ends
and hairpin configurations in the theory and show that
hairpins are described by an Ising-like symmetry-breaking term in the boson
field theory.
\Date{December 1991; Revised February 1992}

\nfig\fI{Hydrodynamic averaging volume surrounding a small region of a
polymer nematic containing many polymer strands.  The average over the
polymer tangents in this volume defines a coarse-grained director
field $\vec n({\bf r},z)$, which then tends to align the individual
polymers which pass through the region.}

\nfig\fII{Conventional short chain nematogens (ellipsoids) connected by
hydrocarbon spacers to make a polymer nematic.}

\nfig\fIII{Isolated polymer in a short chain nematic solvent.
The Goldstone modes of the nematic produce anomalous wandering of the
polymer transverse to the $z$-axis.}

\nfig\fIV{Polymer interacting with director fluctuations in a
{\sl two}-dimensional nematic medium, and described by
its complex co\"ordinate $z$ as a function of arclength $s$.}

\nfig\fIVa{A directed polymer propagating either up or down the $z$-axis,
changing direction whenever a hairpin is present.  Each hairpin
contributes a factor of $w/2$, and each line contributes a free
propagator, $G_0$ to the full propagator $G$.}

\nfig\fV{Contribution to order $h^8$ to the polymer generating function.  Solid
lines are the polymer propagators $G_0({\bf q}_\perpp,q_z)$.  Dashed lines
represent the interaction potential $V(\vec r)$, while the
dotted lines represent interactions induced by the background nematic field.}

\nfig\fVI{Ingredients of the graphical perturbation theory described in the
text.
Figures (a) and (b) represent contributions to the renormalized interpolymer
interaction potential which vanish due to the retarded nature of the
polymer propagator.  The terms shown in (c) are constants which
can be absorbed into a shift in the chemical potential.}

\nfig\fVII{Phase diagram as a function of the boson order parameter and the
chemical potential $\bar\mu$.  The dashed line represents a
contour of constant polymer length.  Typical configurations are
shown at the points $A$ (dilute) and $B$ (dense and entangled).  The
solid curve represents the boson order parameter for $h=0$, and
terminates at a critical point which describes the theory in the
dilute limit.}

\nfig\fVIII{Contours of constant scattering intensity for long chain
ferro- and electrorheological fluids.  The contours are linear
near the origin and surround a maximum on the $q_\perpp$-axis located
approximately at the position of the first Bragg peak of the nearby
hexagonal crystalline phase.  The linear contours near the
origin are rounded by finite chain length effects.}

\nfig\fIX{Contours of constant scattering intensity near the origin
for polymer nematics.  The characteristic square root contours are
rounded off as indicated by the dashed lines unless the polymers are
very long.}

\nfig\fX{Lowest energy solid phase contribution to the correlation function
$G(\rb,\rb';z,z')$, which inserts a flux head and tail into a
crystalline vortex array.  Dashed lines represent a row of vortices
slightly behind the plane of the page.  In (a), a vacancy is created
at ``time'' $z$, which then propagates and is destroyed at time $z'$.
Interstitial
propagation from $z'$ to $z$ is shown in (b).  The energy of the
``string'' defect connecting the head to the tail increases linearly with
the separation in both cases and leads to the exponential decay
of $G(\rb,\rb';z,z')$.}

\nfig\fXI{Lowest order contributions to the perturbation expansion
used in calculating corrections to the zeroth-order parameters in
the theory.  Graphs which are identically $0$ due to the retarded nature of
the polymer propagator are not included.  In this case, the internal lines are
only integrated
over a small spatial momentum shell $e^{-\ell}<q_\perpp<1$, but
are integrated over all ``time''-like momenta $q_z$.
Figures (a) and (b) shows the contributions to the self-energy of the
polymer propagator.  The contributions in (b) are just constants which
we must absorb into a redefinition of the chemical potential, $\bar\mu$.
In (c) we show the graphs which renormalize the four-point coupling
$v$. Figure (d) shows the
corrections to the vertex between the nematic director $\dnb$ and
the polymers.  Figure (e) shows two graphs which contribute to the
four-point coupling but do not vanish identically.  However, their
contribution is irrelevant by power counting.}

\nfig\fXII{
Fixed point flow in $3$ dimensions (d=2).  We show $u$, the effective
hard core repulsion and $\lb^2$.  The flows come into the
origin tangent to the $\lb^2$ axis.  If $u_0<0$ then the flow in
$u_0$ runs off to large negative values, and the system will
go through a gas-liquid phase transition.}

\nfig\fXIII{In these graphs, we represent a hairpin insertion by
a solid square and the insertion of a free end by a solid circle.
Figure (a) shows diagrams which involve closed loops of polymers
interacting with physical polymers.  These graphs do not factor
into a physical part and an unphysical part.  Figure (b) shows
graphs appearing in a theory with free ends which appear to
involve closed loops of polymers, but instead are interacting
with two or more short polymers.  These graphs are indeed physical.}

\nfig\fXIV{Trajectory with varying $\bar\mu$ through the $r_1$-$r_2$ plane,
where $r_1=-\bar\mu+w$ and $r_2=-\bar\mu-w$.  For $w\ne 0$, the transition
from the dilute limit corresponds to an Ising-like phase transition.
The line $r_1=r_2<0$ corresponds to
an XY-like phase.}

\nfig\fXV{Surface on which a vortex loop lies.  The plane, denoted
by hashed lines, is the ``branch disc'' on which the phase $\theta$
jumps discretely by a multiple of $2\pi$.}

\newsec{Introduction}

\subsec{Overview}

The statistical mechanics of directed, interacting lines has
received renewed attention recently.  This problem is, for example,
directly relevant to
the behavior of high-$T_{\tiny C}$ superconductors in a magnetic field\nref
\Ri{D.R.~Nelson, Phys. Rev. Lett. {\bf60} 1973 (1988);
D.R.~Nelson and H.S. Seung, Phys. Rev. B
{\bf 39} 9153 (1989); and D.R. Nelson, J. Stat. Phys., {\bf 57},511 (1989).}
\nref\Rii{For a review, see D.R. Nelson in Proceedings of the Los Alamos
Symposium 1991:
``Phenomenology and Applications of High Temperature Superconductors'',
edited by K. Bedell {\sl et al} (J. Wiley, New York, 1991).}
\refs{\Ri,\Rii}. In these systems, above a critical
external field $H_{{\tiny C}1}$, the magnetic field penetrates the material in
the form of lines, each carrying one quantum of magnetic flux.  The flux
lines can be viewed as ``polymers'' aligned with the direction of the external
field up to thermal fluctuations.
Due to their mutual repulsion, these lines can form various states such as
a triangular solid \ref\Riii{A.A.~Abrikosov, Sov. Phys. JETP {\bf 5} 1174
(1957).}, or isotropic and hexatic entangled fluids\nref\Riv{M.C.~Marchetti and
D.R.~Nelson, Phys. Rev. B {\bf 41} 1910 (1990).} \refs{\Ri,\Riv}.  Glassy
states are also possible, induced either by local disorder\nref\Rv{A.I.~Larkin,
Sov. Phys.
JETP {\bf 31} 784 (1970).}\nref\Rvi{M.P.A.~Fisher, Phys. Rev. Lett.
{\bf 62} 1415 (1989). D.S.~Fisher, M.P.A.~Fisher and D.A.~Huse, Phys. Rev. B,
{\bf 43}, 130 (1991).
E.M.~Chudnovsky Phys. Rev. B {\bf 40} 11355 (1989).} \refs{\Rv,\Rvi} or
simply by very long disentanglement relaxation times\nref\Rvii{S.P.~Obukhov
and M.~Rubinstein, Phys. Rev. Lett. {\bf 65}, 1279 (1990).} \refs{\Ri,\Rvii}.
Detailed calculations for flux liquids are possible by exploiting a
mapping onto the statistical mechanics of boson worldlines in two spatial
and one timelike dimension \Ri .

Many other physical systems consist of extended one-dimensional
objects aligned in one direction\nref\Ra{P.G. de Gennes,{\sl Polymer Liquid
Crystals}, edited by A. Ciferri, W.R. Kringbaum and R.B. Meyer (Academic, New
York, 1982) Chapter 5.}\nref\Rb{R.B. Meyer, {\sl ibid}, Chapter 6.}
\nref\Rc{T. Odijk, Macromolecules {\bf 17},2313 (1986).}\refs{\Ra,\Rb,\Rc}.
Stiff biological macromolecules such as
DNA \ref\Rviii{F.~Livolant and Y.~Bouligand, J. Phys. (Paris) {\bf 47} 1813
(1986), R.~Podgornik, D.C.~Rau and V.A.~Parsegian, Macromolecules {\bf 22}
1780 (1989).}, helical synthetic polypeptides such as {\sl
Poly($\gamma$-benzyl
glutamate}) (PBG)\nref\Rix{H.~Block,
{\sl Poly($\gamma$-Benzyl-L-Glutamate) and other Glutamate Acid Containing
Polymers} (Gordon and Breach, London, 1983).}\nref\Rxi{R.B.~Meyer, F.~Lonberg,
V.~Tarututa, S.~Fraden, S.D.~Lee and
A.J.~Hurd, Disc. Faraday Chem. Soc. {\bf 79} 125 (1985)} \refs{\Rix,\Rxi},
discotic liquid crystals composed of stacks of disk-shaped molecules\nref
\Rxii{S.~Chandrasekhar, B.K.~Sadashiva and K.A.~Suresh, Pramana {\bf 9} 471
(1977).}\nref\Rxiii{Nguyen Huu Tinh, H.~Gasparoux and C.~Destrade, Mol.
Cryst. Liq. Cryst. {\bf 68} 101 (1981). T.K.~Attwood, J.E.~Lyndon and F.~Jones,
Liq. Cryst. {\bf 1} 499 (1986).} \refs{\Rxii,\Rxiii}, and micelles of
amphiphilic
molecules \ref\Rxiv{S.A.~Safran, L.A.~Turkevich and P.~Pincus, J. Phys. Lett.
(Paris) {\bf 45} L69 (1984).} can all form crystalline
columnar phases with in-plane
order, as well as nematics with fluid-like in-plane order.
Although stiffer chains align more easily, some nematic polymer liquid
crystals can also be formed with chains of relatively low rigidity, by
alternating a nematogenic unit with a flexible hydrocarbon spacer
\nref\Rxv{A.~Blumstein, G.~Maret and S.~Villasagar, Macromolecules {\bf 14}
1543 (1981).}\nref\Rxvi{P.G.~de Gennes, C.R. Acad. Sc. Paris {\bf 281} B
101 (1975).}\nref\Rxvii{P.G.~de Gennes, Mol. Cryst. Liq. Cryst. (Letters)
{\bf 102} 95 (1984).}\refs{\Rxv-\Rxvii}. The transition from isotropic
melt to nematic is achieved experimentally by lowering the temperature \ref
\Rxviii{A.~Blumstein, in {\sl Liquid Crystalline Order in Polymers,} Academic
Press, New York ed. (1978). E.M.~Barrall II and J.F.~Johnson, J.~Macromol.
Sci. Rev. Macromol. Chem. {\bf C17} 137 (1979).} or more frequently by
increasing the concentration. Steric repulsion is sufficient to produce
alignment at high enough concentration, although many other interactions
can be present depending on the material: Van der Waals attraction,
electrostatic forces\nref\Rixx{A.~Stroobants, H.N.W.~Lekkerkerker and
Th.~Odijk, Macromolecules {\bf 19} 2232 (1986).}\nref\Rxx{S.~Fraden, G.~Maret,
D.L.D.~Caspar and R.B.~Meyer, Phys. Rev. Lett. {\bf 63} 2068 (1989).}
\refs{\Rixx,\Rxx}, hydration forces, {\sl etc.}.
Ferrofluids \ref\Rd{R.E. Rosenweig, {\sl Ferrohydrodynamics},(Cambridge
University Press, New York, 1985).} and electrorheological fluids \ref\Re{
T.C. Halsey and W. Toor, Phys. Rev. Lett. {\bf 65},2820 (1990), and
references therein.} are also
composed of chains of particles, in this case aligned by external magnetic
or electric fields.

In a recent paper \ref\rexpe{X. Ao, X. Wen and R.B. Meyer, Physica A {\bf 176},
63 (1991).}, X. Ao, X. Wen and R.B. Meyer have presented X-ray scattering
data on PBG which is in many ways strikingly similar to predictions \Ri\
for neutron diffraction by flux lines.  These authors stress an analogy
between polymer configurations and a fictitious ``dynamics'' of two-dimensional
particles moving along the direction of alignment, and show that the
distinctive ``bowtie'' scattering contours change character in the limit
of small momentum transfers.

There are, however, significant differences between these directed polymer-like
objects and flux lines.  Although oriented nematic polymers in a solvent
wander along a preferred axis, just as thermally excited flux lines do, the
average polymer direction represents a spontaneous, rather than externally
imposed, broken symmetry.  Even if the monomer chains are aligned by an
external
electric or magnetic field, the lines are typically of variable length and need
not span the system, unlike flux lines.  Polymer nematics can, moreover, make
relatively low energy hairpin turns, because of the symmetry of the director
field
under $\vec n\rightarrow -\vec n$.  Such ``backtracking'' can usually be
ignored for flux lines \refs{\Ri,\Rii}.

In a recent paper \ref\Rf{P. Le Doussal and D.R. Nelson, Europhys. Lett.
{\bf 15}, 161 (1991).} two of us adapted methods developed for flux lines
to these systems, taking the above differences into account.  We showed
that the boson field theory of \Ri\ becomes applicable to directed
polymer melts upon adding a source term, in analogy with the des Cloiseaux
trick \Ri\ for isotropic polymer solutions.  The
spontaneously broken symmetry of polymer nematics requires, in addition,
coupling the boson order parameter to a massless fluctuating background
director field.

In this work, we describe these calculations in detail, and discuss as well
their validity in the dilute limit.  Calculations for dilute directed polymers
are a straightforward extension of results for flux lines \Ri\ in the case of
electrorheological and ferro-fluids.  The dilute limit is more
interesting, however, when the average polymer direction represents
a spontaneously broken symmetry.  Polymer nematics in an isotropic
solvent will, of course, eventually
crumple into an isotropic phase upon dilution \ref\Rh{For a discussion
of this transition, see L. Balents,
R.D. Kamien, P. Le Doussal, and E. Zaslow, to appear in Journal de Physique,
March 1992.}. An analysis of the transition
to an isotropic polymer melt would take us beyond the scope of this paper.  We
shall, however, consider a dilute collection of polymers aligned by a solvent
which is itself a short chain {\sl nematic}. As pointed out by
de Gennes \Ra , the Goldstone modes associated with the the nematic solvent
already lead to logarithmic anomalies in the wandering of one isolated
polymer.  Additional logarithms appear when interactions are included, and
a full renormalization group analysis is required to sort out the details.
Yet another complication appears when we allow for {\sl hairpins} in the
polymers as they meander through the nematic solvent.  We find in this case
that an {\sl Ising-like} symmetry-breaking term appears in the boson field
theory.

Table I contains a summary ofthe different types of directed polymeric
systems considered in this paper.  Note that flux lines in high temperature
superconductors are the simplest case, because (1) they contain no free ends
and (2) hairpins are highly disfavored by the external magnetic field.
Although hairpins are disfavored by the magnetic and electric fields
necessary to produce aligned chains in ferro- and electrorheological fluids,
free ends are, of course, unavoidable.  The behavior of polymer nematics and
polymers in a nematic solvent is complicated both by the existence of free
ends and hairpins, and because the alignment can be produced by a
``soft'' broken symmetry instead of an external field.

We consider here only {\sl Gaussian} fluctuations about the
state which describes directed polymer melts in the dense hydrodynamic
limit.  While this paper was in preparation, we learned of interesting
work by Toner \ref\rTONER{J. Toner, IBM Preprint (1991).} who introduces
nonlinearities directly into the hydrodynamic theory of polymer nematics.
Toner concludes that these nonlinear terms eventually trigger a breakdown
of hydrodynamics ({\sl i.e.} a singular dependence of hydrodynamic parameters
on wavevector) at sufficiently
long wavelengths.  It would be interesting to see if such a breakdown
also occurred in the more microscopic boson theory developed here.

\subsec{Model}

We shall concentrate on polymer nematics, regarded as directed polymers
interacting with a background nematic field.  By imposing a magnetic field
$H$, or taking the limit of very large Frank constants, we can, if desired,
recover results for directed polymer melts with an
externally imposed direction. We start with a nematic free energy
\eqn\eIi{
F_n = {1\over 2}\int d^2\!r_{\perpp}\int dz
\,\left[K_1(\nabla_{\perpp}\!\cdot\!
\delta {\bf n})^2 + K_2(\nabla_{\perpp}\!\times\!\delta{\bf n})^2 +
K_3(\partial_z
\delta{\bf n})^2 + H(\delta{\bf n})^2\right],}
where the $\{ K_i\}$ are the usual Frank constants for splay, twist
and bend, and $\delta {\bf n}(r) = \left[ \delta n_x({\bf r}_{\perpp},z),
\delta n_y({\bf r}_{\perpp},z)\right]$
is a vector representing a small deviation of
director field $\vec n(\vec r)$ from its average orientation along the
$z$-axis, $\vec n(\vec r)\approx (\delta{\bf n},1)$.  We neglect for
now polymer free ends and hairpin turns, and describe the position
of the {\sl i}th polymer as it traverses the nematic medium by a function
$\vec R_j (z) = ({\bf r}_j(z),z)$.  The $N$ polymer lines interact with
each other and the nematic field via a free energy adapted from \Ra\
\eqn\eIii{\eqalign{
F_p={1\over 2}\kappa\sum_{j=1}^N\int &dz\left({d^2{\bf r}_j\over dz^2}\right)^2
+ {1\over 2}g\sum_{j=1}^N\int dz\left({d{\bf r}_j\over dz} - \delta{\bf n}
({\bf r}_j(z),z)\right)^2\cr
&+{1\over 2}\sum_{i\ne j}^N\int dz V\left(\vert{\bf r}_i(z)-{\bf
r}_j(z)\vert\right)
\cr}}
Here $\kappa$ is the polymer bending rigidity, while $g$ controls the coupling
between the local polymer direction and the nematic matrix.  This coupling
is the only one allowed by rotational invariance, to lowest order in
$d{\bf r}_j/dz$ and
$\delta\bf n$.  The potential $V(\rb)$ represents short range, excluded volume
effects, and can be approximated by
$V(r)\approx V_0{\bf\delta}^2(\rb)$.
The probability of a particular field configuration is proportional to
$\exp (-F/\kb T)$, with $F=F_n+F_p$, and averages are calculated by
integrating over both $\delta{\bf n}(\vec r)$ and
polymer configurations $\{{\bf r}_j(z)\}$.

The simplest physical interpretation of the free energy $F$ is of polymers
aligned by a nematic solvent with Frank constants $\{K_i\}$.  We believe,
however, that $F$ also describes {\sl dense} nematic polymers in an
{\sl isotropic} solvent.  As illustrated in Figure 1, $\vec n(\vec r)$ then
represents a coarse-grained nematic field obtained by averaging over the
polymer tangents in a hydrodynamic averaging volume.  Deviations of the
orientation of any individual polymer from this average direction
are described by the coupling $g$.  This interpretation of $F$ is especially
appropriate for polymers made of nematic molecules connected by flexible
hydrocarbon spacers \refs{\Ra,\Rxvii}\ as in Figure 2.  In this
case $\delta {\bf n}(\vec r)$ describes fluctuations in the orientations of
individual
nematogens, while ${\bf r}_j(z)$ describes how the nematogens are threaded
together by the hydrocarbon spacers.
The bare Frank constants in \eIi\ are then approximately those
of the nematic phase of the unpolymerized nematogens.  In this picture,
we have \Ra\ $\kappa = K_3/\rho_0$, where $\rho_0$ is the
areal density of polymers cutting a constant-$z$ cross section.  Note that
the potential $V(\vec r)$ represents a {\sl scalar} interpolymer
interaction within a constant-$z$ plane.  The coupling $g$ contains
both a scalar interaction, due to the longitudinal modes of the nematic,
and a {\sl vectorial} interaction, due to the transverse nematic modes
(see Section 6).
In going from the bare polymer system to this coarse-grained system, we
could first match the strength of the vectorial interactions with $g$, and
then adjust $V$ so as to get the correct strength of the scalar
interaction.

Our
assumption that polymers interacting with ``nematic background'' field
are equivalent to dense polymer nematics in an isotropic solvent is supported
by the hydrodynamic approach to correlation functions discussed in
Section 5, which gives identical results for these two systems in the limit
of long wavelengths.

The second basic assumption underlying our model calculations is that
we can neglect the term proportional to $\kappa$ in \eIii .  To justify this,
note first that the initial two terms of \eIii\ define a length \Ra\
$\lambda\equiv\sqrt{\kappa /g}$, which for isotropic solvents we
identify with the ``deflection length'' discussed by Odijk \Rc .  The
deflection length is the distance a polymer wanders along $z$ before
it feels the confining effect of its neighbors.  In order that the
polymers order nematically, this length must be less than the polymer
persistence length $\ell_p=\kappa/\kb T$.  On scales larger than
$\lambda$, the coupling to the background nematic field dominates the
bending rigidity and we are justified in neglecting the first term in
\eIii .  If the deflection length is known, we can express the coupling
$g$ in terms of experimental parameters as
\eqn\eIiii{g={K_3\over \rho_0\lambda^2}.}

The equivalence between polymer nematics and polymers in a nematic solvent
does {\sl not} extend to the crumpling transition to a more isotropic
phase which should occur in the dilute limit for polymer nematics.  The
nematic solvent acts like an ordering magnetic field in the latter case, so
that the polymer exhibits nematic order at arbitrary dilution.
We should also mention that
refinements in the model defined by \eIi\ and \eIii\ are required to handle
polymer free ends and hairpins.  These are most easily treated after
rewriting the theory in a second quantized ``boson'' formalism, as discussed
in Section 3.

\subsec{Outline}

In Section 2 we review the wandering of a single polymer in a nematic solvent,
assuming at first that hairpin turns can be neglected.  In an external field,
the polymer just executes a gaussian random walk in the $xy$-plane
as it meanders down the $z$-axis.  When the field is turned off, the
Goldstone modes of the nematic matrix induce logarithmically divergent
``superdiffusive'' behavior, as first pointed out by de Gennes \Ra . We
then generalize the model to include hairpins, using an effective field theory
method introduced by Cardy \ref\ri{J. Cardy, J.Phys. A{\bf 16},L355 (1983).}.
The hairpins cause the polymer to crumple and execute an anisotropic {\sl
three}
dimensional Gaussian random walk.  We also discuss polymers wandering in a
two dimensional nematic solvent.  Below the Kosterlitz-Thouless transition
of the solvent, we find that the polymer exhibits a {\sl continuously
variable} Flory exponent, which is simply related to the decay of order
in the orientations of the solvent molecules.

In Section 3, we show how to solve the model when many interacting aligned
polymers are present via a mapping onto the quantum mechanics of
two-dimensional bosons.  The constraints imposed on the theory by
rotational invariance are discussed in Appendix A.
Adding a source to the boson field theory allows
us to obtain analogous results to polymers of finite length.  In Section 4,
we calculate the polymer density correlation
functions and discuss the renormalized wave vector dependent elastic constants.

In Section
5, we generalize the hydrodynamic approach of de Gennes \ref\Rj{P.G. de Gennes,
J. Phys. (Paris) Lett. {\bf 36L},55 (1975).} and of Selinger and
Bruinsma \ref\Rl{J. Selinger and R. Bruinsma, Phys. Rev. A {\bf 43},2910
(1991).}
to allow
for polymer heads and tails and show that the results agree with the
long wavelength limit of our more microscopic calculations.
In Appendix B, we show explicitly how to derive the hydrodynamic theory
directly from the boson formalism.
We also use the hydrodynamic theory to discuss the physical
meaning of the ``boson'' order parameter used in Sections 3, and to calculate
the elastic energy of a chain end.

The behavior in the dilute limit is discussed in Section 6.  We construct
renormalization
group recursion relations and show how polymer wandering is affected by
both nematic Goldstone modes and interpolymer interactions.  In section 7
we then
introduce hairpins and show that an Ising-like phase transition then
describes the dilute limit.

\newsec{A single chain in a nematic matrix}

\subsec{Three Dimensions: the de Gennes Approximation}

With the problem of the nematic polymer in mind, de Gennes introduced the
following free energy for a single chain without hairpins in a nematic solvent
in the one Frank constant approximation ($K=K_1=K_2=K_3$) \Ra ,
\eqn\eIIi{
F_1 = \half\kappa\int dz\,\left({d^2{\bf r}\over dz^2}\right)^2
+\half g\int dz\,
\left\vert{d{\bf r}\over dz} - \delta{\bf n}({\bf r}(z),z)\right\vert^2
+ \half K\int d^3\!r\,\left\vert\nabla\dnb\right\vert^2
}
The form of the polymer-nematic coupling comes from the small
tipping angle expansion of $\half g\left[1-\left({d\vec R(z)\over dz}
\cdot\vec n\left(\vec R(z)\right)\right)^2\right]$,
which is the only leading order coupling consistent with
rotational invariance and the discrete $\vec n\rightarrow -\vec n$ symmetry.
We use ${d\vec R\over dz}\approx({d\rb\over dz},1)/\sqrt{1+\vert{d\rb\over dz}
\vert^2}$, and $\vec n=(\dnb,1)/\sqrt{1+\vert\dnb\vert^2}$ to find that
\eqn\eIIii{
\left({d\vec R\over dz}\cdot\vec n\right)^2\approx
1-\left\vert{d\rb\over dz}-\dnb\right\vert^2,}
which leads to the coupling displayed in \eIIi .

For a fixed configuration of the polymer, the nematic matrix is distorted and
some elastic energy results (see Figure~3). We wish to compute the
effective free energy of the chain resulting from integrating out the
nematic field in \eIIi .  Although all terms in \eIIi\ are
quadratic, this problem is, in fact, quite nonlinear due to
the appearance of $\rb (z)$ in the argument of $\dnb (\rb (z),z)$ in the
coupling term.
A natural
approximation, implicit in de Gennes' discussion \Ra, is to set
$\rb(z)\simeq {\bf 0}$, and to discuss
only the effect of the fluctuations of the tipping angle $d\rb\over dz$.
This amounts to replacing the $g$-coupling in \eIIi , by:
\eqn\eIIiii{{g\over 2}\int dz\,\left\vert {d\rb\over dz}-\dnb(\rb(z),z)
\right\vert^2
\rightarrow{g\over2}\int dz\,\left\vert {d\rb\over dz}-\dnb({\bf
0},z)\right\vert^2,}
where $\dnb(0,z)$ is the value of the distortion of the nematic
field on the line $\rb ={\bf 0}$. One expects
this approximation to be better for large $K$.
A perturbation expansion in $K^{-1}$ of the original problem \eIIi\
shows that for $d\le 3$, divergent integrals appear, and one must use the
renormalization group to obtain correct results. We defer
a systematic treatment of this problem to Section 6 and discuss here only
the de Gennes approximation,
which can be expected to be qualitatively correct in $d=3$.

The fastest way to compute the effective free energy of the polymer is to
note that \eIIi\ modified by \eIIiii\ can
be simplified by the change of variable (with unit Jacobian) ${\bf\eta}(z)=
d\rb/dz-\dnb(0,z)$. Upon neglecting the bending energy term, we find
that $\rb(z)$ is the solution of
\eqn\eIIiv{{d\rb\over dz}=\dnb({\bf 0},z)+\eta(z)\;,}
where $\eta(z)$ is a Gaussian white noise of variance $\kb T/g$. The wandering
of the chain is then given by integrating \eIIiv\ over $z$, squaring and
averaging over $\dnb({\bf 0},z)$,
\eqn\eIIv{\langle\vert \rb(L)-\rb(0)\vert ^2\rangle=
{2\kb T\over g}\,L+
\int_0^L \int_0^L dz\,dz'\;
\langle\dnb({\bf 0},z)\cdot\dnb({\bf 0},z')\rangle_{F_\n}\;.}
Due to the Goldstone mode associated to the rotational invariance of the
nematic free energy $F_\n$, the director correlation function decreases like
$1/\vert z-z'\vert $ at large separations.  Upon substituting the full Frank
energy \eIi\ for the one-Frank-constant term in \eIIi\ we find (upon setting
$H=0$),
\eqn\eIIvi{\langle\dnb({\bf 0},z)\cdot\dnb({\bf 0},0)\rangle =
{\kb T\over4\pi}\,\lt[{1\over K_1}+{1\over K_2}\rt]\,{1\over z}\;.}
The long range correlations in the medium then implies via \eIIv\
``hyperdiffusion'' of
the chain in the transverse direction:
\eqn\eIIvii{\langle\vert \rb(L)-\rb(0)\vert
^2\rangle\mathrel{\mathop\sim_{L\rightarrow\infty}}
{2\kb T\over g}\,L+
{\kb T(K_1+K_2)\over 2\pi K_1K_2}\;L\,\ln\,(L/a)\;,}
where $a$ is a microscopic cutoff.
The effective ``diffusion
constant'' (defined by $\langle\vert
\rb(L)-\rb(0)\vert^2\rangle=4D(L)L\,$)
is finite for finite $L$, but diverges logarithmically as $L\rightarrow\infty$.
\eqn\eIIviii{D(L)\equiv{\kb T\over 2g} + {\kb T(K_1+K_2)\over 8\pi K_1K_2}
\ln(L/a)}
For $K_1$=$K_2$=$K$, this is the result obtained by de Gennes by
explicitly carrying out the integration over the nematic field \Ra .
The renormalization group treatment of Section 6 (which includes
interpolymer interactions) leads to the same result with, however, a factor
of two difference in the coefficient of the logarithm.

The nematic solvent is not quite able to
produce a finite renormalized diffusion constant for infinitely long
polymers. The polymer is still aligned with the $z$-axis on large length
scales, however, since $\sqrt{\langle\vert \rb(L)-\rb(0)\vert^2\rangle} \ll
L$.
It is easy to show
that the bending energy,
which has been neglected in these calculations, is irrelevant
for a single chain without hairpins.

It is interesting to note that, despite its annealed character, this
problem is very similar to the problem of random walks in quenched random
disorder for which long range correlations are known to modify
diffusion \ref\Rxxix{J.P. Bouchaud and A. Georges, Physics Reports {\bf
195},127
(1990); P. Le Doussal and J. Machta, Phys. Rev. B{\bf 40}, 9427 (1989).}.

\subsec{Two Dimensions}

It is interesting to ask what the configurations $\vec r(s)$
are of a chain in a two
dimensional nematic. Although more difficult than its three-dimensional
counterpart, an
experiment on a two-dimensional surface might be possible:  Imagine
a long polymer chain with $N$ monomers adsorbed at an air
water interface which is also covered with a tilted monolayer Langmuir-Blodgett
film.  The projection of the tilted hydrocarbon chains on the the plane
of the interface plays the role of a director $\vec n$, {\sl without} the
inversion symmetry $\vec n\rightarrow -\vec n$. We parameterize the director by
$\vec n=(\cos\phi,\sin\phi)$. Thermal fluctuations
destroy the tilt order above $T_{\tiny\rm C}$
via a Kosterlitz-Thouless \ref\Rxxx{D.R. Nelson, in {\sl Phase Transitions
and Critical Phenomena}, Vol. 7, edited by C. Domb and
J.L. Lebowitz (Academic, New York, 1983), and references therein.}
vortex unbinding transition.
Even below $T_{\tiny\rm C}$, however, there is no privileged
direction in the tilt field
since a broken continuous symmetry is impossible in two dimensions for systems
with short range interactions. Thus, one expects the polymer to crumple for
$T<T_{\tiny\rm C}$, and one can ask
for the wandering exponent $\nu$ governing the mean end to end distance,
$R\sim N^\nu$, where $N$ is the number of monomers. Note that $\nu=\half$
in the three-dimensional problem discussed above.
The problem for $d=2$ is
nontrivial because below $T_{\tiny C}$, the nematic field has long range
correlations
decaying with a continuously varying exponent:
\eqn\eIIx{\langle \vec n({\bf x})\cdot\vec n({\bf x}') \rangle =
\langle e^{i\phi({\bf x})}e^{-i\phi({\bf x}')}\rangle\sim\vert {\bf x}-{\bf
x}'\vert ^{-\eta(T)}\;.}
The exponent $\eta(T)$ varies from $\eta=0$ at $T=0$ to $\eta={1\over 4}$ at
the transition \Rxxx .
There are {\sl a priori} two Frank constants $K_1$ and $K_3$ in two-dimensions,
but it
has been shown \ref\Rxxxi{D.R. Nelson and R.A. Pelcovits, Phys. Rev. B {bf 16}
2191 (1977).}\
that
at large length scales two dimensional nematics become isotropic
($K_1=K_3=K$) and are described at long wavelengths
by the usual XY model free energy ${K\over2}\int dx\,(\nabla\phi)^2$.
For $T<T_{\tiny\rm C}$ one has $\eta=\eta(T)={\kb T\over
2\pi K}$. Above $T_{\tiny\rm C}$ the correlations decay
exponentially in \eIIx, and then one expects that the wandering exponent
of the polymer is the pure self-avoiding random walk value $\nu=3/4$ (or
$\nu=1/2$
for an ideal chain).

A simple random walk argument gives an
interesting prediction for the exponent $\nu$ below $T_{\tiny\rm C}$.
In the infinite $g$ limit
one expects that the polymer be totally aligned with the local field.
\eqn\eIIxi{{dz\over ds}=e^{i\phi(z(s))}\;,}
where we have used the complex notation for the position $z(s)=x_1(s)+ix_2(s)$,
$s$
being the arclength along the polymer. Upon integrating over $s$ we find
\eqn\eIIxii{\langle\vert z(s)-z(0)\vert ^2\rangle_{z(0),\phi}=
\int_0^s \int_0^s du\,du'
\langle e^{i[\phi(z(u))-\phi(z(u'))]} \rangle_{z(0),\phi}}
where the average is over both initial conditions $z(0)$ for the polymers
and over the smoothly varying director angle $\phi(z)$.
We now neglect the reaction of the polymer on the
nematic field and replace the correlation in the right
hand side of \eIIxii\ by
the correlation of the unperturbed nematic field \eIIx . This procedure
amounts to replacing the problem by the one of determining the wandering
of the tangent curves of
a pure nematic (see Figure 4). We are led to an integral equation, namely
\eqn\eIIax{\langle\vert z(s)-z(0)\vert^2\rangle_{z(0)} =
\int_0^s\int_0^sdu\,du'\left\langle{{\rm const}\over\vert z(u)-z(u')\vert^\eta}
\right\rangle_{z(0)}}
Upon assuming that $\vert z(s)-z(0)\vert$ scales as $s^\nu$, we find that
the self consistent value of $\nu$ is
\eqn\eIIxiii{\nu(T)={2\over2+\eta(T)}\;.}
Our results coincide with those of Flory theory applied to random walks in
quenched random environments, which has been argued to be exact for
divergenceless flows\nref\Rxxxii{P. Le Doussal, to be published.}
\refs{\Rxxix,\Rxxxii}.

According to Eq. \eIIxiii\ $\nu$ is always close to $\nu=1$, continuously
decreasing with increasing temperature from $\nu(T=0) =1$ to $\nu(T=
T_{\tiny\rm C}) = 8/9$.  Note that $\nu$ always exceeds $3/4$,
the value of a self-avoiding random walk in two dimensions.  Our neglect
of self avoidance for $T<T_{\tiny\rm C}$ is thus self-consistent
because of the relatively small number of self-intersections which occur
for $\nu>3/4$.

\subsec{Field Theoretic Treatment of Hairpins}

In this section we study a {\sl single}
polymer embedded in a {\sl rigid} nematic matrix, and show that hairpins
induce an Ising-like crumpled state.  A field theory due to Cardy \ri\ is used
to
describe the polymer and introduce hairpins, providing a
simplified example of the boson field theory used in later sections.  We show
explicitly that hairpins cause a single directed polymer to crumple into
a more isotropic configuration at long length scales, as suggested in
\Ra .  Although crumpling {\sl always} occurs in a single polymer, we
show in Section 7 that interactions induce an Ising-like
transition from a dilute crumpled state to directed state stabilized by
interactions when many polymers
are present.

Consider the following model free energy for a single polymer in
a nematic matrix $\vec n$,
\eqn\epierreone{
F=\int ds\,\left[{\kappa\over 2}\left({d\vec T\over ds}\right)^2 - {g\over 2}
\left(\vec T\cdot\vec n\right)^2\right]}
where $\vec T={d\vec R(s)\over ds}$ is the three dimensional tangent vector to
the
polymer, and $s$ is the arclength.  The first term is the bending
energy, while the second represents the lowest order coupling to the
background nematic compatible with overall rotational symmetry.  Higher order
terms in $\vec T$ are possible, but are inessential for the following
discussion.  Note that $\vert\vec T\vert =1$.

Suppose the fluctuations in $\vec n$ are suppressed by imposing a large
magnetic field, so that $\vec n=\hat z$.  Then one has
\eqn\epierretwo{
F=\int ds\,\left[{\kappa\over 2}\left({d\vec T\over ds}\right)^2 - {g\over 2}
(T_z)^2\right]}
which is exactly a one dimensional Heisenberg model with a quadratic
Ising-like anisotropy.  As a function of $T_z$, there are two minima
for $T_z=\pm 1$.  If $g$ is very large, one can expand perturbatively
around each minimum.  Because this is a one dimensional problem, the
symmetry is in fact restored by tunneling events between the two minima,
which correspond to hairpins.  $T_z$ thus
plays the role of an Ising variable and hairpins are analogous to an Ising
domain wall along the one dimensional chain.  The tunneling probability from
one minimum to the next is the energy of a hairpin occurring over a
distance $R$, with energy $\epsilon_h\approx\min_R\left({\kappa\over
2R}+{gR\over 2}\right)
\sim(g\kappa)^{1/2}$, and
the radius of the hairpin is $R_h\sim(\kappa/g)^{1/2}$ which coincides
with de Gennes more elaborate calculation \Ra .  Note that the size of a
hairpin is also of the order of the Odijk deflection length $\lambda$ \Rc .
The typical distance between two hairpins along the chain is thus
$l\approx ae^{\epsilon_h/\kb T}$, which can be very large.  The statistics
of the hairpins is analogous to the statistics of one dimensional
Ising domain walls (with Ising coupling $J\sim\epsilon_h$).  This
analogy implies that the length scale $a$ is just the Odijk deflection
length.  Note that a stretching force applied to the endpoint of the chain
is equivalent to an Ising magnetic field, since $\int ds\,hT_z = h(z(L)-z(0))$.
The Ising magnetization is the analog of the total size of the polymer along
$\hat z$.  The elastic modulus $G$ for Hooke's law along $\hat z$ is thus
given by the Ising susceptibility, $G\sim (1/T)e^{J/\kb T}$ as $T\rightarrow
0$.

The effect of director fluctuations will be discussed in Section 7.  Here,
we continue to neglect them and implement the above ideas with an
Ising-like field theory.  As noted by Cardy the propagator of a {\sl single}
directed polymer of variable length along $\hat z$ with $2$ transverse
dimensions
can be written as the correlation function of a quadratic action, which
we write directly in the continuum limit \ri ,
\eqn\epierrethree{
G(\rb,{\bf 0};z,0) = {1\over Z}\int {\cal D}\psi{\cal D}\psi^*\,
\psi^*({\bf 0},0)\psi(\rb,z)e^{-S_0}}
where
\eqn\epierrefour{
S_0=\int dzd^2\!r\,\psi^*\left(\partial_z - D\nablab_\perpp^2 -\bar\mu\right)
\psi}
with $D=\kb T/2g$.  Here $G(\rb,{\bf 0};z,0)=
\sum W_{\tiny N}(\rb,{\bf 0};z,0)e^{-\bar\mu N}$ and
$W_{\tiny N}(\rb,{\bf 0};z,0)$ is the total weight for
directed walk of $N$ steps to begin at $({\bf 0},0)$ and end
at $(\rb,z)$. The chemical potential $\bar\mu$ must be
adjusted to give the correct polymer size.
The use of a retarded propagator only and the
neglect of self-interactions if hairpins are excluded has
been justified by Cardy.
The critical point of this quadratic theory
is $\bar\mu\rightarrow 0^-$ corresponding to infinite length
chains.  The propagator in Fourier space is then
\eqn\epierrefive{
G({\bf q}_\perpp,q_z) = {1\over -iq_z + Dq_\perpp^2 -\bar\mu}\equiv
G_0({\bf q}_\perpp,q_z).}

Now we discuss a single directed polymer with hairpins but {\sl without}
self avoidance (see Figure 5).  One can introduce hairpins by
simply adding to the action \epierrefour\ the term
\eqn\epierree{S=S_0+{w\over 2}\int dzd^2\!r\,\left[\psi^2 +
\left(\psi^*\right)^2
\right]}
which allows for ``pair creation''.  Let us consider the new term as
a perturbation and expand in $w$ the above correlation function \epierrethree .
One then generates diagrams as in Figure 5, with a factor $w$ per hairpin
and $G_0({\bf q}_\perpp,q_z)$ per solid line.  All closed loops are canceled
by the normalization factor $Z$, so the $n\rightarrow 0$ trick is
unnecessary.  The term of order
$w^{2n}$ is the sum of all
possible ways to go from $({\bf 0},0)$ to $(\rb,z)$ with
$2n$ hairpins -- the odd terms vanish.
The analogy of hairpins with kinks in an Ising-like
system shows that we should take $w\propto e^{\epsilon_h/\kb T}$, with
$\epsilon_h\sim\sqrt{g\kappa}$.  If one follows the connected line from
$({\bf 0},0)$ the factor associated with the part which represents
backward propagation ({\sl e.g.}, after an odd number of hairpins) is
actually $G^*_0({\bf q}_\perpp,q_z)$.  Thus the sum of
all these diagrams is
\eqn\epierreol{\eqalign{
G&=G_0({\bf q}_\perpp,q_z)\left[
1+\left({w\over 2}\right)^2G_0G_0^* + \left({w\over 2}\right)^4(G_0G_0^*)^2
+\ldots\right]\cr
&={G_0\over 1-{w^2\over 4}G_0G_0^*}\cr}}
Thus the propagator is now
\eqn\epierreprop{
G={iq_z + Dq_\perpp^2-\bar\mu\over
q_z^2 + (Dq_\perpp^2-\bar\mu)^2-{w^2\over 4}}}
The critical point has been shifted and is now at $\bar\mu=-w/2$. This
corresponds
physically to the fact that allowing hairpins adds an extra entropy in
the system, and the proliferation of paths occurs earlier when
$\bar\mu$ is raised from $-\infty$.  To find the asymptotic
large distance behavior, we set $\bar\mu=-w/2 +\delta\bar\mu$ and expand
for small $\delta\bar\mu$, ${\bf q}_\perpp$, and $q_z$.  Then we have
\eqn\epierreagain{
G\sim{1/2\over (q_z/w)^2 + Dq_\perpp^2 -\delta\bar\mu}}
Note that complete propagator is actually twice the above result
(more precisely it is $G+G^*$) because one has now to allow for graphs where
the first
propagator in the above series goes backwards. This
is the propagator for an (anisotropic) Gaussian random
walk in $d$-dimensions. The polymer size now
scales as $\sqrt{N}$ both along $\hat z$ and along $\rb$, but
with an anisotropic radius of gyration tensor.  The {\sl
directed} nature of the walk has disappeared.
\newsec{Field Theory for a Liquid of Chains with a Nematic Background}

In this section we use the analogy with the statistical mechanics of
two-dimensional
bosons to describe the properties of chains. The details of this analogy
are discussed in\nref\rpolp{D.R. Nelson and P. Le Doussal, Phys. Rev. B{\bf
42},
10113 (1990).} \refs{\Ri,\rpolp}, so we will indicate here only the differences
with the case of the flux lines. For a fixed configuration of the nematic
solvent, the analogy works as before, except that now the bosons are also
interacting with an external ``time'' and space dependent field. The average
over configurations of the nematic field has to be carried out at the end.
We initially discuss the theory for chains which span the system.  Internal
free ends are then introduced by adding a source to the boson coherent state
field theory.  The field theory with the free ends is then used to calculate
some important correlation functions in Section 4.

\subsec{Mapping onto two-dimensional bosons}

We start from the following partition function for the $N$ chains in a nematic
field discussed in the introduction
\eqn\eIIIi{\eqalign{&Z_N[\dnb]={1\over N!}\int\prod_{i=1}^N {\cal D}\rb_i(z)\cr
&\quad e^{\displaystyle{
\lt(-{1\over \kb T}\int_0^L dz\lt[\sum_{i=1}^N{g\over2}
\lt({d\rb_i\over dz}-\dnb(\rb(z),z)\rt)^2+\sum_{i<j}^N
V(\rb_i(z)-\rb_j(z))\rt]\rt)}}\;,\cr}}
where we have omitted the bending term
and restricted our attention to
interactions between lines through an equal ``time''
potential $V(\rb_i(z)-\rb_j(z))$.
The quantity which we are ultimately interested in is the average:
\eqn\eIIIii{Z_N=\int {\cal D}\dnb(r,z)\,Z_N[\dnb]\,
\exp\lt[-{F_n\over \kb T}\rt]\;,}
where $F_n$ is given by \eIi .

\nref\rIIIa{J.W. Negele and J. Orland, {\sl Quantum Many-Particle Systems}
(Addison-Wesley, New York, 1988) chaps. 1 and 2.}\nref\rIIIb
{V.N. Popov, {\sl Functional Integrals and Collective Excitations}
(Cambridge University Press, New York, 1987).}
Well known transformations \refs{\rIIIa,\rIIIb}\ using transfer matrices allow
to write
$Z_N[\dnb]$ in the Hamiltonian form (first quantization),
\eqn\eIIIiv{\eqalign{Z_N[\delta\n]={1\over N!}\int
&d^2\!r_1\ldots d^2\!r_Nd^2\!r'_1\ldots d^2\!r'_N\cr
&\lt\langle\rb'_1\ldots \rb'_N
\lt\vert {\bf T}\exp\lt\{-\int dz\,{\ham(z)\over\kb T}\rt\}\rt\vert
\rb_1\ldots\rb_N\rt\rangle ,\cr} }
where $\vert \rb_1\ldots\rb_N\rangle$ and $\vert \rb'_1\ldots\rb'_N\rangle$
are states corresponding to the entry and exit points of the polymers \Ri\ and
the time ordering operator $\bf T$ is necessary {\sl a priori} since $\ham$ is
$z$ dependent. $\ham$ can be deduced from
the Lagrangian
\eqn\eIIIabc{
\lag(\rb_i,\dot\rb_i,z)={g\over2}\sum_i[\dot\rb_i - \dnb(\rb_i,z)]^2 +
\sum_{i<j} V(\rb_i-\rb_j)}
according to the usual rules
of Euclidean quantum mechanics \rIIIa
\eqn\eIIIv{\ham(\pb_k,\rb_k,z)=\lag+i\sum_j\pb_j\cdot\rb_j\;,}
where $\rb_j$ has been eliminated using $\pb_j=i{\pt \lag\over\pt\dot\rb_j}$.
The
quantization rule is then $\pb_j=-i\kb T\nablab_j$.
We define $\nablab$ so that it operates only within the plane
perpendicular to $\hat z$.  The quantity $\bf r$ will always mean
a vector perpendicular to $\hat z$. Following these rules, we
obtain the Hamiltonian for a fixed configuration of $\dnb$:
\eqn\eIIIvi{\ham={-(\kb T)^2\over2g}\sum_j\nabla_j^2+\kb
T\sum_j{1\over2}(\nabla_j
\dnb(\rb_j,z)+\dnb(\rb_j,z)\nabla_j)+{1\over2}\sum_{i\ne j}V(\rb_i-\rb_j)\;,}
where we take a symmetrical ordering \rIIIa .

A coherent state representation can now be developed in analogy to the
treatment of flux lines in \Ri\ , and one obtains for the grand canonical
partition function:
\eqn\eIIIviii{Z_{\rm gr}\equiv\sum_{N=0}^\infty e^{L\mu N/kT}Z_N=
\int {\cal D}\psi^*(\rb,z)\,{\cal D}\psi(\rb,z){\cal
D}\dnb(\rb,z)\,\exp-S[\psi^*,\psi,\dnb]\;,}
where $\psi(\rb,z)$ is a complex boson field.  The boson action $S$
reads
\eqn\eIIIix{
S=\eqalign{&\int_0^L dz\int d^2\!r\,\left[
\eqalign{
&\psi^*(\rb,z)\lt({\pt\over\pt z}-
D\nablab_\perp^2-\bar\mu\rt)\psi(\rb,z)\cr
&+{1\over 2}\left[\psi^*(\rb,z)\nablab_\perp\psi(\rb,z)
-\psi(\rb,z)\nablab_\perp\psi^*(\rb,z))\right]\cdot\dnb(\rb,z)\cr
&+{1\over2}\int d\rb'\,\bar v(\rb-\rb')\vert \psi(\rb,z)\vert ^2\vert
\psi(\rb',z)\vert ^2\cr
}\right]\cr
&\qquad\qquad
+{F_n[\dnb]\over \kb T}\cr}\;.}
Similar manipulations show that the density of flux lines is
\eqn\eIIInext{
\rho(\rb,z)=\vert\psi(\rb,z)\vert^2}
where
$D=\kb T/2g$, $\bar\mu =\mu/\kb T$ and $\bar v =V/\kb T$.
In Section 4, we shall calculate correlations in the density
$\rho(\rb,z) = \sum_{i=1}^N\delta[\rb - \rb_i(z)]$
may be calculated via the identification \eIIInext .  In the mean field
approximation, we have $\rho_0 =\langle\vert\psi\vert^2\rangle=\bar\mu/\bar v$.
The field theory embodied in \eIIIix\ differs from the action for flux lines
\Ri\ only in the coupling of the boson ``current'' $\psi^*\nablab_\perp\psi -
\psi\nablab_\perp\psi^*$ to the director field.  As discussed in Appendix~A,
rotational invariance of the original ``Lagrangian'' model \eIIIi\ ({\sl i.e.}
``Lorentz invariance'' of the fictitious bosons) forces the
coefficient of
$(\psi^*\nablab_{\!\perpp}\psi-\psi\nablab_{\!\perpp}\psi^*)\cdot\dnb$ to
be exactly half that of $\psi^*\partial_z\psi$ in the second
quantized coherent state representation.

\subsec{Effects of Finite Chain Length}

To allow for polymers of finite length, which start and stop in the
interior of the sample, we add a source term to \eIIIix
\eqn\eIIIni{S\rightarrow S-h\int dz\int d^2\!r\,[\psi+\psi^*].}
Upon replacing $\psi$ and $\psi^*$ by their mean field average value
$\sqrt{\rho_0}$, we see that the quantity $h\sqrt{\rho_0}$ is the probability
per unit
area and per unit ``time'' $z$ of starting or terminating a
polymer.  Since an average of one polymer will thread a cross-sectional
area $\rho_0^{-1}$ perpendicular to $z$, the typical polymer length
associated with this Poisson-like process is $\ell=\sqrt{\rho_0}/h$

A more formal proof of the equivalence of bosons with a source
to the statistical mechanics of directed polymers of finite length can
be constructed along the lines taken for isotropic polymer
melts\nref\rnewc{J. des Cloiseaux, J. Phys. (Paris) {\bf 36}, 281 (1975);
see also R.G. Petschek and P. Pfeuty, Phys. Rev. Lett. {\bf 58}, 1096 (1987).}
\nref\rnewd{P.G. de Gennes, {\sl Scaling Concepts in Polymer Physics}
(Cornell University Press, Ithaca, 1970).} \refs{\rnewc,\rnewd} .
We first expand the partition function associated with \eIIIix\ in $h$,
$\bar v$, and the nonlinear coupling to the director field.  The Fourier
transformed propagator in the resulting Feynman diagrams (see Fig. 5) is
\eqn\eIIIten{
G_0({\bf q}_\perpp,q_z) = {1\over -iq_z + Dq_\perpp^2},}
or in real space
\eqn\eIIIeleven{
G_0(\rb,z) \equiv \langle\psi(\rb,z)\psi^*({\bf 0},0)\rangle_0
=\Theta(z) {1\over 4\pi Dz}e^{-r^2/4Dz},}
where $\Theta(z)$ is the step function.
The expectation value in \eIIIeleven\ is taken with respect to the Gaussian
action
\eqn\eIIItwe{
S_0=\int_0^Ldz\int d^2\!r\,\left[\psi^*(\partial_z -
D\nablab_\perpp^2)\psi\right]}
Equation \eIIIeleven\ is identical to the propagator used in the
polymeric description of flux lines presented in \Ri .  We can
think of $\psi^*({\bf 0},0)$ as creating a polymer at $({\bf 0},0)$ and
$\psi(\rb,z)$ as destroying this polymer at $(\rb,z)$.  Although
this produces the usual diffusive random walk propagator for $z>0$,
$G_0(\rb,z)=0$ when $z<0$, showing that propagation
backwards in the time-like variable $z$ is impossible.

The grand canonical partition function may now be expressed formally as
\eqn\egrand{
Z_{\rm gr} = \sum_{N_m=0}^\infty\sum_{p=0}^\infty {\cal Z}_{N_m}^p(D,\bar v,
K_i)e^{\bar\mu L_m}h^{2p},}
where ${\cal Z}_{N_m}^p(D,\bar v,K_i)$ is the partition function for $N_m$
monomers distributed among $p$
polymers and $L_m\propto N_m$ is the {\sl total} length along
$\hat z$ occupied by the $N_m$ monomers.  Note that $h^2$ plays the role
of a polymer fugacity, while $e^{\mu}$ controls the density of monomers.
Figure 6 shows a typical contribution to ${\cal Z}_{N_m}^p$ of order
$h^8$.  The solid lines are the polymer propagator $G_0(q_\perpp,q_z)$
discussed above.  Dashed lines represent the interaction potential $V(\rb)$,
while the dotted lines signify interactions induced by the background nematic
field.
The retarded nature of the propagators insures that large numbers of
``unphysical'' graphs disappear.  The graph shown in Figure 7a, for
example, with zero momenta on its external legs, is proportional to
\eqn\esix{
\int {dq_zd^2\!{q_\perpp}\over (2\pi)^3}{\vert V({{\bf q}_\perpp})\vert^2\over
(-iq_z+D{q_\perpp}^2)^2},}
which vanishes because both poles in the $q_z$ integration are on the
same side of the real axis.  The graph shown in Figure 7b does not
contribute for similar reasons. The graphs shown in Figure 7c are constants
which can be absorbed into a redefinition of the chemical potential.

By using the expansion \egrand , we can easily show that the average
polymer length is given by
\eqn\eseven{
\langle L_m\rangle={\partial\over\partial\bar\mu}\ln Z_{\rm gr}=
{\sum_{N_m=0}^\infty\sum_{p=0}^\infty L_m{\cal Z}_{N_m}^p(D,\bar v,
K_i)e^{\bar\mu L_m}h^{2p}\over Z_{\rm gr}}}
while the average number of polymers is
\eqn\eeight{
\langle p\rangle=h^2{\partial\over \partial (h^2)}\ln Z_{\rm gr}
={\sum_{N_m=0}^\infty\sum_{p=0}^\infty p{\cal Z}_{N_m}^p(D,\bar v,
K_i)e^{\bar\mu L_m}h^{2p}\over Z_{\rm gr}}}
We can now calculate the typical polymer
length
\eqn\eaxx{\eqalign{\ell&\equiv\langle L_m\rangle/\langle p\rangle\cr
&={\langle\vert\psi(\rb,z)\vert^2\rangle\over {h\over
2}\langle\psi(\rb,z)+\psi^*(\rb,z)\rangle}\cr
}}
directly from the mean field approximation to the partition function \eIIIviii
{}.
We assume for simplicity the contact potential $V(\rb)=\kb T\bar v
\delta^2(\rb)$.
Upon making the substitution \eIIIni\ in \eIIIviii\ and setting $\psi(\rb,z)$
to a constant value $\psi=\psi^*=\psi_0=\sqrt{\rho_0}$, we have
\eqn\eaxxi{
\ln Z_{\rm gr} = -\Omega\min_{\psi_0}\left\{-\bar\mu\psi_0^2+{1\over 2}\bar
v\psi_0^4
-2h\psi_0\right\}}
where $\Omega$ is the (three dimensional) volume.  There are two limiting cases
to consider.  When $\bar\mu\gg 0$, we assume an ordered state only
slightly perturbed by the small source field $h$.  As we shall see, this
means restricting our attention to very long chains.  The minimum of \eaxxi\ is
then given to lowest order in $h$ by
\eqn\eaxxii{\psi_0\approx\sqrt{\bar\mu\over \bar v}\left[1+\sqrt{\bar
v\over\bar \mu^3}
{h\over 2}\right]}
while the partition function is
\eqn\eaxxiii{
Z_{\rm gr}=\exp\left[\Omega\left({\bar\mu^2\over 2\bar v} +
2h\sqrt{\bar\mu\over\bar v}
\right)\right]}
We then find from \eseven\ and \eeight\ that
\eqn\eaxxxiv{\langle L_m\rangle\approx\psi_0^2\Omega}
and
\eqn\eaxxxv{\langle p\rangle\approx \psi_0h\Omega}
so that a typical length is
\eqn\eaxxxvi{\ell\approx\psi_0/h ,}
as also follows directly from the mean field approximation.
Note that this length diverges as $h\rightarrow 0$, and agrees
with the estimate made at the beginning of this section.

If $\bar\mu\roughly{<} 0$, the polymers are dilute, we are in the disordered
phase of the model.  We assume that polymers follow an imposed external
direction, as in electrorheological fluids, so that we
need not worry about the crumpling transition discussed in the Introduction.
One can then neglect $\bar v$ and find that the order parameter is
\eqn\eaxxxvii{\psi_0\approx h/\vert\bar\mu\vert}
while the grand partition function is
\eqn\eaxxxviii{{\cal Z}_{\rm
gr}\approx\exp\left[\Omega\left({h^2\over\vert\bar\mu\vert}
\right)\right]}
The total length of polymer is now
\eqn\eaxxix{\langle L_m\rangle={h^2\over\bar\mu^2}\Omega}
while the average number is
\eqn\eaxxx{\langle p\rangle={h^2\over\vert\bar\mu\vert}\Omega}
A typical polymer size is thus
\eqn\eabugs{\ell\approx{1\over\vert\bar\mu\vert}\qquad\qquad\qquad(h=0^+)}
which increase as the transition is approached from negative $\mu$.

Experiments are usually done varying the polymer concentration with a
fixed distribution of polymer sizes, and hence, fixed $\ell$.  This
produces the trajectory shown in Figure 8 on the phase diagram plotted as
a function of $-\mu$ and $\psi_0$.  Typical polymer configurations at two
points on this phase diagram are indicated.  Note that the model corresponds
to a {\sl distribution} of chain lengths instead of a monodisperse
sample.  This is also a feature of the des Cloiseaux model of isotropic
polymer melts, for which it is possible to draw a very similar
phase diagram \refs{\rnewc,\rnewd}.  A broad
distribution of chain lengths is probably a good approximation for
electrorheological fluids and ferrofluids, in which the chains are
constantly breaking and reforming.  The difference between polydisperse
and monodisperse samples is not, in any event, expected to be important
when the average chain length is large \rnewd .

\newsec{Results for Correlation Functions}

We now assume that the polymers are dense and entangled, {\sl i.e.}
$\sqrt{D\ell}\gg \rho_0^{-1/2}$.  Correlations for small $h$ can
then be calculated as in \refs{\Ri,\rpolp}, by expanding about
the mean field order parameter $\psi_0=\sqrt{\rho_0}\approx\sqrt{\bar\mu/\bar
v}[1+{\cal O}(h)]$.

\subsec{Gaussian Form for the Action}

Let us write the complex field $\psi(\rb,z)$ in terms of the density
$\rho(\rb,z)$ and the phase $\theta(\rb,z)$ as:
\eqn\eIIIxi{\psi(\rb,z)=\rho(\rb,z)^{1/2}\exp(i\theta(\rb,z))\;.}
The measure takes the simple form, up to unimportant constant factors:
\eqn\eIIIxii{{\cal D}\psi^*(\rb,z){\cal D}\psi(\rb,z)={\cal D}\rho(\rb,z)
{\cal D}\theta(\rb,z)\;.}
To simplify the discussion, let us again consider a contact potential
$V(\rb)=\V_0
\delta^2(\rb)$, and set $\bar v=V_0/\kb T$. The action then takes the form in
the new variables:
\eqn\eIIIxiii{\eqalign{
S=&\int dzd^2\!r\lt[\eqalign{&i\rho\pt_z\theta
+{D\over 4\rho}(\nablab_\perpp\rho)^2+D\rho(\nablab_\perpp\theta)^2
+i\rho\nablab_\perpp\theta\cdot\dnb\cr&-\bar\mu\rho+{1\over2}\bar v\rho^2
-2h\rho^{1/2}\cos\theta\cr}\rt]\cr
&+{F_n[\dnb]\over \kb T}.\cr}}
Upon setting $\rho(\rb,z)=\rho_0+\delta\rho(\rb,z)$,
where $\rho_0=\psi_0^2$ is given by \eaxxii ,
and expanding to quadratic order the fluctuations $\delta\rho$ and
$\theta$, the action becomes, up to total derivatives and constants,
\eqn\eflourfour
{S=\int dzd^2\!r\left[
\eqalign{&i\delta\rho\partial_z\theta +{D\over 4\rho_0}
(\nablab_\perpp\delta\rho)^2+D\rho_0(\nablab_\perpp\theta)^2+i\rho_0
\nablab_\perpp\theta\cdot\dnb\cr
&+{1\over 2}\bar v(\delta\rho)^2 +h\rho_0^{1/2}\theta^2 +{1\over
2}h\rho_0^{-3/2}
(\delta\rho)^2\cr}\right]}
Note that the term involving $\dnb$ can be rewritten as $-i\rho_0\theta
(\nablab_\perpp\cdot\dnb)$, so that only the longitudinal part of $\dnb$
couples to
the polymer degrees of freedom.

Upon expanding the fields in Fourier modes
\eqn\efixo{\theta(\rb,z)=(\Omega)^{-1/2}
\sum_{{\bf q}_\perpp,q_z } e^{izq_z +i{\bf q_\perpp}\cdot\rb}\theta({\bf
q}_\perpp,q_z ),}
with similar expansions for $\delta\rho$ and $\dnb$, one obtains
the action:
\eqn\eIIIxv{S={1\over2}\sum_{{\bf q}_\perp,q_z }
X^+({\bf q}_\perpp,q_z )G^{-1}({\bf q}_\perpp,q_z )X({\bf q}_\perpp,q_z )\;,}
where $X\equiv(\theta,\delta\rho,\delta\n_x,\delta\n_y)$ and, (taking $H=0$
in \eIi\ for simplicity),
\eqn\eIIIxvi{\eqalign{&G^{-1}=\cr&\,\lt[\matrix{
\left({{\eqalign{\scriptstyle2&\scriptstyle
D\rho_0q_\perpp^2\cr&\scriptstyle+2h\rho_0^{1/2}\cr}}}\right)
&\scriptstyle -q_z  &\scriptstyle -\rho_0q_{x} &\scriptstyle -\rho_0q_{y}\cr\cr
\scriptstyle q_z  &
\left({{\eqalign{&\scriptstyle{D\over2\rho_0}q_\perpp^2+\bar v\cr
&\scriptstyle\quad
+ h\rho_0^{-3/2}\cr}}}\right) &\scriptstyle 0 &\scriptstyle 0 \cr\cr
\scriptstyle\rho_0q_{x} &\scriptstyle 0 &
{\scriptstyle(K_1q_{x}^2+K_2q_{y}^2+K_3q_z^2)\over \scriptstyle\kb T} &
{\scriptstyle K_1-K_2\over \scriptstyle\kb T}
\scriptstyle q_{x}q_{y}\cr\cr
\scriptstyle \rho_0q_{y} & \scriptstyle 0 & {\scriptstyle
K_1-K_2\over\scriptstyle \kb T}\scriptstyle q_{x}q_{y}
& {\scriptstyle(K_1q_{y}^2+K_2q_{x}^2+K_3q_z^2)\over \scriptstyle\kb T}\cr\cr
}\rt]\cr}}
It is now straightforward to calculate any desired correlation function, by
inverting this $4\times 4$ matrix.

\subsec{Discussion of Correlations}

The structure function
\eqn\efoursix{
S({\bf q}_\perpp,q_z) = \langle\vert\delta\rho({\bf
q}_\perpp,q_z)\vert^2\rangle}
may be written
\eqn\efourseven{
S({\bf q}_\perpp,q_z)=\rho_0{\rho_0q_\perpp^2+2\bar K(\vec
q)(\ell^{-1}+Dq_\perpp^2)\over
\bar K(\vec q)[q_z^2+\bar\epsilon^2(q_\perpp)]+{1\over
2}\rho_0q_\perpp^2\bar\epsilon^2
(q_\perpp)/(\ell^{-1}+Dq_\perpp^2)}}
where
\eqn\efouregiht{\bar K(\vec q)={K_1q_\perpp^2+K_3q_z^2\over \kb T}}
we have set $\rho_0^{1/2}/h=\ell$, and
\eqn\efournien{\bar\epsilon^2(q_\perpp)=(\ell^{-1}+Dq_\perpp^2)(\ell^{-1}
+Dq_\perpp^2+2\rho_0v).}
To obtain results for arbitrary in-plane pair potentials $V(\rb)$, let
$\bar v\rightarrow\hat V({\bf q}_\perpp)/\kb T$, where $\hat V({\bf q}_\perpp)$
is the
Fourier transform of $V(\rb)$.  Unlike its flux line counterpart,
the ``Bogoliubov spectrum'' \efournien , has a gap as $q_\perpp\rightarrow 0$,
due to the finite polymer lengths.  Upon evaluating $\langle\dn_i(\vec
q)\dn_j(-\vec q)\rangle$,
we determine the renormalized Frank constant via the identification
\eqn\efourten{
\langle\dn_i(\vec q)\dn_j(-\vec q)\rangle = {1\over K_2^{\tiny R}q_\perpp^2
+K_3^{\tiny R}q_z^2}P^{\tiny T}_{ij} + {1\over K_1^{\tiny R}q_\perpp^2
+K_3^{\tiny R}q_z^2}P^{\tiny L}_{ij}}
$K_2$ and $K_3$ are unrenormalized, while
\eqn\efourtena{K_1^{\tiny R}({\bf q}_\perpp,q_z)=K_1 + {1\over 2}\rho_0\kb T
{\bar\epsilon^2(q_\perpp)\over
q_z^2+\bar\epsilon^2(q_\perpp)}{1\over(\ell^{-1}+
Dq_\perpp^2)}.}

Upon taking the limit $\bar K(\vec q)\rightarrow\infty$, which suppresses all
fluctuations in the director field, \efourseven\ can be applied to polymer
melts with an externally imposed average direction.  In the limit of
small wavevectors, the structure function takes the simple form,
\eqn\efoureleven{
S({\bf q}_\perpp,q_z)=\kb T{\rho_0^2q_\perpp^2 +K/G\over Bq_\perpp^2 + Kq_z^2
+G^{-1}KB
\rho_0^{-2}}}
where $B=\rho_0^2V_0+{\cal O}(\ell^{-1})$ is a two dimensional bulk
modulus, $K=\rho_0g$ is a tilt modulus and $G$ is related to the
average length via $G=\ell\kb T/2\rho_0$.  When $\ell\rightarrow\infty$ the
structure function contours are straight lines passing through the origin, with
no scattering along the line $q_\perpp=0$ \refs{\Rvi,\Rvii} .  Scattering
reappears at $q_\perpp=0$ for finite $\ell$, however.  Fits of \efoureleven\
to scattering data should allow a simple direct determination of the average
chain length $\ell$ and the important parameters $B$ and $K$.

The small wave vector limit of \efourseven\ when the Frank constants are
finite has a different form
\eqn\efourtwelve{
S({\bf q}_\perpp,q_z)=\kb T{\rho_0^2q_\perpp^2+(K_1q_\perpp^2+K_3q_z^2)/G\over
Bq_\perpp^2+(B/G\rho_0^2+q_z^2)(K_1q_\perpp^2+K_3q_z^2)}.}
When $G=\ell\kb T/2\rho_0\rightarrow\infty$, \efourtwelve\ reduces to the
prediction \Rl\ of a hydrodynamic theory due to de Gennes \Rj .
Finite length effects distort the contours near
the origin, however, and are likely to be quite important in fitting
real scattering date.  Fits to \efourtwelve\ should lead to direct
experimental determination of the Frank constant, the bulk
modulus $B$ and the mean polymer length.  Upon taking the limit $\vec
q\rightarrow 0$
in \efourtena\ we find the dependence on the renormalized splay
elastic constant $K_1^{\tiny R}$ on the polymer length,
\eqn\efourthirteen{
K_1^{\tiny R}=K_1+{1\over 2}\kb T\ell\rho_0 ,}
in agreement with a prediction of R.B. Meyer \Rb , but in disagreement with
a suggestion by de Gennes \Ra .

\newsec{Hydrodynamic Treatment of Correlations}

The hydrodynamic description of isotropic liquids of atoms or small molecules
has been understood for many years \ref\rbigf{M.E. Fisher, Rep. Prog. Physics
{\bf 30}, 615 (1967).}.  The long wavelength density fluctuations are
Gaussian, and static correlation functions are described by Ornstein-Zernike
theory \ref\roz{L.S. Ornstein and F. Zernike, Proc. Acad. Sci. Amst. {\bf 17},
793 (1914).}.  We generalize here an analogous theory of liquids of
oriented lines in three dimensions.  The basic concepts were first
discussed by P.G. de Gennes and R.B. Meyer in the context of polymer nematic
liquid crystals over a decade ago \refs{\Ra,\Rb}.  These ideas were
recently used to determine the form of the liquid crystal structure function
near the origin of reciprocal space by Selinger and Bruinsma \Rl .
The hydrodynamic theory can in fact be derived directly from the boson
representation, as shown in Appendix B.
We show here how a finite density of chain lengths can be incorporated in a
natural
way for ferro- and electrorheological fluids, as well as for polymer nematics.
The results agrees in all cases with the long wavelength limit of those in
Section 4.
We also use the hydrodynamic theory to show that there are essentially
no differences at long wavelengths between polymers in a nematic solvent
and dense polymer nematics in an isotropic solvent.  Some
of our conclusions were reviewed recently in \ref\rlines{
D.R. Nelson, Physica A {\bf 177}, 220 (1991).}.

Hydrodynamics will, in addition, allow us to discuss the physical meaning of
the boson order parameter introduced in Section 3.  To this end, we calculate
the energy of an isolated free end in a ferro- or electrorheological fluid.
Such a calculation has already been carried out for polymer nematics by
Selinger
and Bruinsma \ref\rsb{J. Selinger and R. Bruinsma, UCLA preprint.}.

\subsec{Ferro- and Electrorheological Fluids}

We first assume for simplicity that the chains of magnetic or electric
dipole particles
span the system along the $\hat z$-axis.  The effects of
finite chain length will be put in later.  The basic hydrodynamic
fields are now an areal particle density
\eqn\efiveone{
\rho_{\rm mic}(\rb,z)=\sum_{j=1}^N\delta^2[\rb-\rb_j(z)]}
and a ``tangent'' field in the plane perpendicular to
$\hat z$,
\eqn\efivetwo{
{\bf t}_{\rm mic}(\rb,z)=\sum_{j=1}^N{d\rb_j(z)\over dz}\delta^2[
\rb-\rb_j(z)].}
We coarse grain these microscopic fields to obtain smoothed
density and tangent field $\rho(\rb,z)$ and ${\bf t}(\rb,z)$.

We now expand the free energy of the liquid to quadratic order in the
density deviation $\delta\rho(\rb,z)=\rho(\rb,z)-\rho_0$ and
in ${\bf t}(\rb,z)$
\eqn\efivethree{
F={1\over 2\rho_0^2}\int d^2\!r dz\,\left[B(\delta\rho)^2 + K\vert{\bf
t}\vert^2\right].}
The parameter $B$ is a bulk modulus for areal compressions and dilations
perpendicular to the $z$-axis, while $K$ is the modulus for tilting
lines away from the direction of the applied field.  Because we
are dealing with {\sl lines}, and not simply oriented anisotropic particles,
\efivethree\ must be supplemented with an ``equation of continuity,''
\eqn\efivefour{\partial_z\delta\rho +\nablab_\perp\!\cdot\!{\bf t}=0,}
which reflects the fact that vortex lines cannot stop or start inside the
medium. Correlation functions can be calculated by assuming that the
probability of a particular line configuration is proportional to
$\exp(-F/\kb T)$, and imposing the constraint \efivefour\ on the
statistical mechanics.

We now modify \efivethree\ to allow for chains which stop and start inside the
medium.  If the chains are long and entangled, we can treat the chain
heads and tails as independent ideal gases, following similar
ideas for polymer nematics by R.B. Meyer \Rb . The free energy \efivethree\
is now replaced by an expansion of the form
\eqn\efivefive{
F={1\over2}\int d^2\!r dz\left[B\left({\delta\rho\over\rho_0}\right)^2
+K\left({{\bf t}\over \rho_0}\right)^2 + G(\partial_z\delta\rho+\nablab_\perp
\!\cdot\!{\bf t})^2\right]}
which differs from \efivethree\ only in a term proportional to the square
of the constraint displayed in \efivefour . Although other terms proportional
to gradients of $\delta\rho$ and $\bf t$ can appear,  we have
kept only those couplings which dominate in the limit of
long chains.  We can now treat $\delta\rho$
and $\bf t$ as independent fields and then take the limit $G\rightarrow\infty$
to impose the constraint. The coupling $G$ is finite, however, when the chains
are of finite length.  To determine the value of $G$ in this case, note
that the chain heads and tails act as sources and sinks on the right
hand side of the conservation law \efivefour . It follows that \Rb
\eqn\efivesix{\partial_z\delta\rho+\nablab_\perp\!\cdot\!{\bf t}=\rho_{\tiny H}
-\rho_{\tiny T},}
when finite densities of chain heads $\rho_{\tiny H}(\rb,z)$ and
chain tails $\rho_{\tiny T}(\rb,z)$ are present.  If the heads
and tails are treated as two noninteracting ideal gases, a term of the form
\eqn\efiveseven{
\delta F={1\over 2}G\int d^2\!r dz\,(\rho_{\tiny H}-\rho_{\tiny T})^2}
should appear in the free energy.  The coefficient is just the concentration
susceptibility for an ideal binary mixture, which is well known to be
\ref\rmq{As follows, e.g., from the solution of Problem 5-13, in D.A.
McQuarrie,
{\sl Statistical Mechanics} (Harper and Row, New York, 1976).}
\eqn\efiveeight{G={\kb T\over\langle\rho_{\tiny H}\rangle+\langle\rho_{\tiny T}
\rangle}}
where $\langle\rho_{\tiny H}\rangle$ and $\langle\rho_{\tiny T}\rangle$ are the
average concentrations of heads and tails, respectively.  Now, each chain
contributes both a head and a tail, $\langle\rho_{\tiny H}\rangle=\langle
\rho_{\tiny T}\rangle=\rho_{\rm chain}$, and the
three-dimensional density of chains is $\rho_{\rm chain}=\rho_0/\ell$, where
$\ell$ is a typical chain length. It follows that
\eqn\efivenine{G={\ell\kb T\over 2\rho_0},}
which diverges as $\ell\rightarrow\infty$.  It is easy to check that the
structure function which results from this hydrodynamic treatment agrees
with the result \efoureleven\ of the more microscopic boson calculations
of Section 3 in the long wavelength limit.

Figure 9 shows the scattering contours expected for dense aligned ferro-
or electrorheological fluids.  The maxima along the $q_\perpp$-axis occur
approximately at the position of the first reciprocal lattice vector in
the nearby crystalline phase, and are {\sl not} accounted for by the
hydrodynamic theory.  As discussed in \Ri , the half-width at
half-maximum along $q_z$ for fixed $q_\perpp$ controls
the decay of density fluctuations along the $z$-axis.  Hydrodynamics,
(in agreement with the boson theory) determines via \efoureleven\ the linear
form of the contours near the origin.  The rounding of these contours
(indicated by the dashed lines) due to the finite polymer length is one of the
principal predictions of \Rf\ and this paper.  When $\ell\rightarrow\infty$,
the contours remain linear as $q_\perpp,q_z\rightarrow 0$ and the scattering
vanishes along the $q_z$-axis, as in the case of flux lines \Ri .

\subsec{Polymer Nematics}

The hydrodynamic treatment of infinitely long polymer nematics is due
originally to de Gennes \Rj , and was applied to correlation
functions by Selinger and Bruinsma \Rl .  We first assume an
isotropic solvent.  The hydrodynamic variables are now the areal polymer
density \efiveone\ and a fluctuating nematic deviation field
$\dnb(\rb,z)$ attached to the polymers.  We again allow
for a dilute concentration of chain ends by writing the hydrodynamic
free energy as
\eqn\efiveten{F={1\over 2}\int d^2\!r
dz\left[B\left({\delta\rho\over\rho_0}\right)
+G(\partial_z\delta\rho + \rho_0\nablab_\perp\!\cdot\!\dnb)^2\right] +
F_n[\dnb]}
where $F_n$ is given by \eIi , and $G$ must again be given by
\efivenine\ in the dilute limit.  When $G\rightarrow\infty$, we recover
the constraint,
\eqn\efiveeleven{\partial_z\delta\rho+\rho_0\nablab_\perp\!\cdot\!\dnb=0}
For finite $G$, one again finds agreement with correlations calculated from
\efiveten\ and the hydrodynamic limit \efourtwelve\ of the result \efourseven\
obtained from the microscopic boson theory.

Figure 10 shows the hydrodynamic predictions for scattering off the polymers
in the limit of small momentum transfer.  The structure for larger
$q_\perpp\sim
\rho_0^{-1/2}$ should be similar to that shown near the peaks in Figure 9.
\rexpe .
For infinitely long polymers, the scattering vanishes along the
$q_z$-axis, and the contours take the form $q_z\propto q_\perpp^{1/2}$ \Rl .
The boson and hydrodynamic theory prediction \efourtwelve\ leads to the
rounding indicated by the dashed lines in Figure 10, an effect which is
likely to be quite important in fitting real experimental data.

As discussed in \rlines , hydrodynamics also makes interesting predictions
about freeze fracture experiments on directed polymer melts.  It can be shown,
in particular, that density fluctuations measured in a fracture plane
perpendicular to $\hat z$ provide a precise signature that one is dealing
with a liquid of lines rather than a liquid of points.  Under favorable
circumstances, it is even possible to determine a typical polymer length
from such measurements.

We can also treat
polymers in a {\sl nematic} solvent using hydrodynamics, and check
that there are no significant differences with conventional polymer nematics
in the long wavelength limit.  The (unpolymerized) nematic solvent
will now be described by the free energy \eIi\ and we introduce
coarse grained polymer variables as in \efiveone\ and \efivetwo .  The
hydrodynamic free energy (as derived explicitly in Appendix B) is then
\eqn\efivetwelve{
F'={1\over 2}\int d^2\!r dz\,\left[B\left({\delta\rho\over\rho_0}\right)^2
+g\vert{\bf t}-\rho_0\dnb\vert^2 + G(\partial_z\delta\rho +
\nablab_\perp\!\cdot\!{\bf t})^2
\right] + F_n[\dnb]}
where the coupling proportional to $g$ reflects
the analogous coupling in \eIIi .  We now integrate
out the solvent degrees of freedom, which leads to an effective free
energy
\eqn\efivethirteen{e^{-F'_{\rm eff}/\kb T}=\int{\cal D}\dnb\,e^{-F'/\kb T}}
given by
\eqn\efivefourteen{F'_{\rm eff} = {1\over 2}\int d^2\!r dz\,
\left[\left({\delta\rho\over\rho_0}\right)^2+G(\partial_z\delta\rho +
\nablab_\perp\!\cdot\!{\bf t})^2
\right] + F_n'[{\bf t}]}
where
\eqn\efivefifteen{\eqalign{
&F_n'[{\bf t}]=\cr
&\int {d^d\!q_\perpp\over (2\pi )^d}{dq_z\over 2\pi}
\left[\eqalign{
&g{K_1q_\perpp^2+K_3q_z^2\over
K_1q_\perpp^2+K_3q_z^2+g\rho_0^2}{q_\perpp^iq_\perpp^j\over q_\perpp^2}\cr
&+g{K_2q_\perpp^2+K_3q_z^2\over K_2q_\perpp^2+K_3q_z^2+g\rho_0^2}
{\delta^{ij}-q_\perpp^iq_\perpp^j\over q_\perpp^2}\cr}\right]t_i(\vec
q)t_j(-\vec q)\cr} }
At long wavelengths, the coefficients of the
longitudinal and transverse projectors in \efivefifteen\ simplify,
and $F_n'$ for polymers in a nematic solvent
reduces to the $F_n$ appropriate for conventional polymer nematics.
The two systems are indeed equivalent from the point of
view of long wavelength polymer correlation functions.

\subsec{Defect Energies and the Boson Order Parameter}

Hydrodynamics also allows us to better understand the boson order
parameter used in Sections 3 and 4.  For an analogous
discussion for flux lines in high $T_{\rm\tiny C}$ superconductors,
see \Rii .  Note that the representation \eIIIxi\ only makes sense if there is
``phase coherence'' in the equivalent ``boson'' system.
Consider the correlation function
\eqn\eVone{
G(\rb,\rb';z,z')=\langle\psi(\rb,z)\psi^*(\rb',z')\rangle}
where $\psi(\rb,z)$ and $\psi^*(\rb',z')$ are, respectively,
creation operators for polymer heads and tails.  We assume that
the polymers are long and entangled, so that $h=0$ in \eIIIni .
Phase coherence means long range order in $G(\rb,\rb';z,z')$,
\eqn\eVtwo{
\lim_{\vert\rb-\rb'\vert\rightarrow\infty}\quad\lim_{\vert
z-z'\vert\rightarrow\infty}
G(\rb,\rb';z,z')={\rm const}>0.}
To understand what this long range order means, consider first the
behavior of \eVone\ in a hexagonal {\sl crystal} of directed polymers, with
the strands again aligned with the $z$-axis.  The composite operator
in \eVone\ creates an extra line at $(\rb',z')$, ({\sl i.e.}, a column of
interstitials in the solid), and destroys an existing line at $(\rb,z)$,
creating a column of vacancies.  As shown in Figure 11, the lowest energy
configuration is then a line of vacancies (for $z'>z$) or interstitials
(for $z'<z$) connecting the two points with an energy $\sigma s$ proportional
to the length $s$ of this ``string.''  The string tension $\sigma$ will
depend on the angle $\theta$ this line makes with the $z$-axis.  It follows
that the correlation function \eVone\ decays exponentially to {\sl zero} ({\sl
i.e.},
like $\exp(-\sigma(\theta)s/\kb T)$) for large
separations in this crystalline phase.  In a directed polymer {\sl melt}, on
the other hand, the concept of vacancy and interstitial lines has no
meaning, and the string tension $\sigma$ will vanish for large $s$, implying
long range order in $G(\rb,\rb';z,z')$.

Long range order in the boson order parameter $\psi(\rb,z)$ means that
\eqn\eVthree{
\langle\psi(\rb,z)\rangle=\langle\psi^*(\rb,z)\rangle\propto e^{-E^*/\kb T}>0.}
Here, $E^*$ is the energy of an isolated polymer head or tail. The hydrodynamic
theory provides a transparent demonstration that unlike in hexagonal
crystals, this energy is finite.
This energy has already been calculated by Selinger and Bruinsma for polymer
nematics \rsb , so we concentrate here on the hydrodynamics for ferro-
and electrorheological fluids.  The constraint \efivefour\ applies everywhere
away from
an isolated head or tail, and is conveniently implemented by expressing
$\delta\rho$ and $\bf t$ in terms of a two component ``displacement field''
${\bf u}(\rb,z)$,
\eqna\ephon{$$\eqalignno{
\delta\rho&=-\rho_0\nablab_\perpp\cdot{\bf u}&\ephon a\cr
{\bf t}&=\rho_0{\partial{\bf u}\over\partial z}&\ephon b\cr
}$$}
following a similar trick by Taratura and Meyer \ref\rTM{
V.G. Taratura and R.B. Meyer, Liquid Crystals {\bf 2},373 (1987).}\ for
polymer nematics.  The free energy becomes
\eqn\eVXX{
F=\half\int dzd^2\!r\,\left[ B(\nablab_\perpp\cdot{\bf u})^2 + K(\partial_z{\bf
u})^2
\right] .}

The extremal equations associated with \eVXX\ for an isolated polymer
head at the origin are
\eqn\eVXXI{
B\nablab_\perpp(\nablab_\perpp\cdot{\bf u}) + K\partial^2_z{\bf u}
=B\rho_0^{-1}\Theta(z)\nablab_\perpp\delta^2(\rb),}
where $\rho_0$ is the average in plane polymer density and $\Theta(z)$ is
the step function, $\Theta(z)=1, z>0$ and $\Theta(z)=0, z<0$.  See
\ref\rtone{J. Toner and D. R. Nelson, Phys. Rev. B{\bf 23}, 316 (1981).}
for a discussion of the source term for the closely related problem of a
dislocation in a two-dimensional smectic liquid crystal.  Here, the
source represents the absence of a line along the positive $z$-axis.
In a crystal, this line would be a string of vacancies, and its energy would
be infinite, due to the disruption of the crystalline order
parameter in the vicinity of the line.  In a liquid, however, the
vacancy free energy vanishes and we need only
consider the long range strain field associated with \eVXXI .
Upon solving \eVXXI\ in Fourier space, we find
\eqn\eVXXII{
{\bf u}({\bf q}_\perpp,q_z)={-(B/\rho_0){\bf q}_\perpp\over
q_z(Bq_\perpp^2+Kq_z^2)}}
from which leads to an explicit expression for the tilt field
\eqn\eVXXIII{
{\bf t}=\rho_0\partial_z{\bf u} = {B^{1/2}K\over 4\pi}{\rb\over
(Kr^2+Bz^2)^{3/2}}.}
The density perturbation associated with the free end is
\eqn\eVXXIV{
\delta\rho=-\rho_0(\nablab_\perpp\cdot{\bf u})=
{-B^{3/2}K\over 4\pi}{z\over (Kr^2+Bz^2)^{3/2}}.}
Since both these ``strains'' fall off like $1/({\rm distance})^2$ far
from the origin, the integrated strain energy in \eVthree\ is
indeed finite, {\sl i.e.},
\eqn\eVXXV{E^*<\infty,}
where we have included a microscopic short range contribution in the defect
energy $E^*$.

The analysis for polymer nematics \rsb\ is more complicated but also
leads to the conclusion that the energy of an isolated head or tail is finite,
consistent with a nonzero value of the boson order parameter.  The
strains in \eVXXIII\ and \eVXXIV\ resemble those expected for
a magnetic monopole in an anisotropic medium.  Indeed, \efivefour\  is
just the condition of ``no magnetic monopoles,'' $\vec\nabla\cdot\vec b=0$,if
we identify $\delta\rho$ with $b_z$ and $\bf t$ with ${\bf b}_\perpp$,
while \efivethree\ is just the magnetic field energy of a
medium with an anisotropic permeability.  The energy of isolated
magnetic monopoles is indeed expected to be finite in three dimensions,
except when the field lines form an Abrikosov flux lattice \Rii .

\newsec{Renormalization Group Analysis of the Dilute Limit}

The analysis in the previous sections applies when the chains
are dense and entangled.  To treat the dilute limit, we
perturb about the limit in which $\langle\psi\rangle\approx 0$.  To this end,
we again consider the (coherent state) functional integral representation of
the partition function:
\eqn\Z{
{\cal Z}_{gr} = \int {\cal D}\psi{\cal D}\psi^*{\cal D}\dnb\,\exp
-S[\psi,\psi^*,\dnb],}
and break the action into three parts,
\eqn\ethreeparts{\eqalign{
S[\psi,\psi^*,\dnb]=&\int d^d\!r\int dz\,\left[\psi^*\left(\partial_z
-D\nablab_\perp^2 -\bar\mu\right)\psi + v\vert\psi\vert^4\right]\cr
&+{\lambda\over 2}\int d^d\!r\int dz\,\dnb\cdot\left(\psi^*\nablab_\perp\psi
-\psi\nablab_\perp\psi^*\right)\cr
&+F_n[\dnb]/\kb T\cr}.}
This form of the action is just \eIIIix , specialized to the case
$\bar v(\rb)=v\delta^d(\rb)$ and with a new
parameter $\lambda$ to help organize perturbation
theory in the coupling between the polymer and nematic degrees of freedom.
As in in \Ri , we impose a cutoff $\Lambda$ on the perpendicular wavevectors
${\bf q}_\perpp$, of order the inverse range of the interaction.
For now we neglect the possibility of free ends and do not introduce a
source term.
It is convenient to consider $d+1$ dimensional directed polymers with
$d$-directions perpendicular to the average direction.  We take the
limit $d=2$ at the end of the calculation.

We can now attempt to expand quantities of physical interest in the nonlinear
couplings $v$ and $\lambda$.  Consider, for example, the propagator,
\eqn\esixtwo{
G(\rb,z)=\langle\psi(\rb,z)\psi^*({\bf 0},0)\rangle,}
or, in Fourier space
\eqn\esixthree{
G({\bf q}_\perpp,q_z)=\langle\vert\psi({\bf q}_\perpp,q_z)\vert^2\rangle.}
This can be written as
\eqn\esixfour{
G({\bf q}_\perpp,q_z) = {1\over -iq_z+Dq_\perpp^2-\bar\mu+\Sigma({\bf
q}_\perpp,q_z)},}
where, to lowest, non-trivial order in $v$ and $\lambda$, the self
energy graphs shown in Figure 12 lead to
\eqn\eSF{\eqalign{\Sigma&({\bf q}_{\perpp},q_z ) =
-\bint\,{dk_z\over 2\pi}\!\bint {d^d\!k_{\perpp}\over (2\pi )^d}\cr
&\left\{\eqalign{&{\lambda^2\over
4}{(2q_{\perpp}-k_{\perpp})^i(2q_{\perpp}-k_{\perpp})^j
\over -i(q_z-k_z )
+ D(q_{\perpp}-k_{\perpp})^2 - \bar\mu}\left(\prop{k_{\perpp}}{k_z}\right)\cr
&\qquad - 2v{1\over -ik_z + Dk_{\perpp}^2 - \bar\mu}\cr}\right\}\cr}}
The first term comes from the nematic-polymer interaction, while
the second term arises from the polymer-polymer interaction.
This self interaction of a single polymer is not present in the original
microscopic model \eIii , and arises here because of the nonuniversal
cutoff dependence (in $q_z$) of the
Feynman path integrals and their representation as a coherent state functional
integral \Ri .  We interpret the integral over $q_z$ in such a way that
its imaginary part vanishes.  The real part is well-defined and simply
gives a constant renormalization of the chemical potential.
After integrating over $k_z$ and evaluating \eSF\ in the zero-frequency,
long wavelength regime we find
\eqn\eSFp{\eqalign{\Sigma&({\bf q}_{\perpp},0) =
-{\lambda^2\over 4}\bint {d^d\!k_{\perpp}\over (2\pi
)^d}{k_{\perpp}^2+(4/d)q_{\perpp}^2
\over
2\displaystyle{\sqrt{K_1K_3}}\vert k_{\perpp}\vert
(\sqrt{K_1/K_3}\vert k_{\perpp}\vert + Dk_{\perpp}^2-\bar\mu)}\cr
&-{\lambda^2\over 4}\bint {d^d\!k_{\perpp}\over (2\pi
)^d}{4(1-1/d)q_{\perpp}^2\over
2\displaystyle{\sqrt{K_2K_3}}\vert k_{\perpp}\vert (\sqrt{K_2/K_3}\vert
k_{\perpp}
\vert
+ Dk_{\perpp}^2-\bar\mu)}+v\bint {d^d\!k_{\perpp}\over (2\pi )^d}\cr}}
Although the correction to $\bar\mu$ in \esixfour\ is well-behaved, the
renormalization of $D$
represented by the coefficient of $q_\perpp^2$  diverges for $d\le 2$, when
$\bar\mu\approx 0$.
This infrared divergence suggests that renormalization group
techniques be employed to study the system for $d\le 2$.  Our approach
follows
\ref\rFSN{D. Forster, D.R. Nelson, and M.J. Stephen, Phys. Rev. A{\bf 16}:732
(1977).},
which is a variation of the dynamic renormalization
group of
\ref\rHH{B.I. Halperin, P.C. Hohenberg, and S.K. Ma,
Phys. Rev. Lett.{\bf 29}:1548 (1972).}.
First we integrate out a momentum shell in $k_\perp$
from $\Lambda e^{-\ell}$ to $\Lambda$, but integrate freely over $k_z$.
We then rescale our variables so that the ultraviolet
cutoff is held fixed. After rescaling $\psi$ and $\dnb$ accordingly,
we are left with the same theory but with different coupling constants.
When this procedure is iterated, $\lambda$ and $v$ are driven toward a fixed
point which describes the universal long wavelength behavior in the dilute
limit.

\subsec{Momentum Shell Integration}

We must first integrate out the transverse momentum in the
range $\Lambda e^{-\ell}<q_\perpp<\Lambda$. This can be done straightforwardly
by expanding the functional integral \Z\ in $v$ and $\lambda$.  The expansion
can
be represented diagrammatically as in Figure 12.  Care must be taken to
account for all possible contractions of the operators in the
expansion.  The symmetry factors can be found in the usual way for
Wick expansions.  It is important to note that diagrams
renormalize the remaining low momentum modes, and are not
simply expectation values.
The diagrammatic rules may be summarized as follows
\point For each line, assign a momentum $k_i$
while conserving energy and momentum at the
vertices.
\point For each polymer line include a factor of $\displaystyle{1\over
-ik_z + Dk_{\perpp}^2 - \bar\mu}$\ .  The sign of $k_z$ is determined
by the direction of the line.
\point For each nematic line include a factor of
$\displaystyle{\prop{k_{\perpp}}{k_z}}$\ .
\point If a vertex joins four polymer lines, include a factor
of $2v$.
\point If a vertex joins two polymer lines and one nematic line, include
a factor of $\lambda\over 2$ and the sum of the incoming and outgoing polymer
momentum.
\point Divide by the symmetry factor.  The symmetry factor is the
product of the number
of ways of permuting the vertices and the number of ways
of permuting the lines while keeping the contractions the same.
\point Integrate over all momenta $q_z$, but
only integrating the transverse momenta
from $\Lambda e^{-\ell}$ to $\Lambda$.

Upon carrying out this procedure, we arrive at the following relations for the
intermediate values of the coupling constants:
\eqna\eINT{$$\eqalignno{
D'&=D\left( 1 + {\lambda^2\over D}
\left[{K_1(d-1) + K_2\over K_1K_2}\right] {A_d\Lambda^{d-2}\over 2d}
{(1-e^{-(d-2)\ell })\over d-2}\right) &\eINT a\cr
\lambda'&=\lambda
&\eINT b\cr
v'&=v - \left( {v^2\over 4D} - {\lambda^2v\over 4K_1D} +
{\lambda^4\over 16K_1^2D} \right) {A_d(1-e^{-(d-2)\ell})\over d-2}
&\eINT c\cr
K_i'&=K_i,\quad i=1,2,3&\eINT d\cr
\bar\mu'&=\bar\mu - \left({v\over D} + {\lambda^2\over 8K_1D}\right)
{A_dD\Lambda^d\over d}\left( 1-e^{-d\ell}\right)&\eINT e\cr}$$}
where $A_d = 2/[\Gamma ({d\over 2})(4\pi )^{d/2}]$.
In evaluating the integrals we have ignored terms of two types.
If a term diverges as $\ell\rightarrow\infty$
in a smaller dimension than the most divergent term, it
is irrelevant by power counting, further,
there are terms which appear to diverge
in a higher dimension, but they are all higher order in the
external momenta.  When we rescale we must rescale these momenta.  Doing
so will render these terms irrelevant in the renormalized and
rescaled theory.  The intermediate values of $D$, $\bar\mu$, and
the Frank constants reflect all the contributions to them at one loop order,
while the expressions for $\lambda$ and $v$
represent only the most relevant contribution to their values.

These intermediate coupling constants have not yet been rescaled.  We rescale
lengths by $L'=Le^{-\ell}$ and times by $T'=Te^{-\int_0^\ell
\gamma(\ell')d\ell'}$.
We now set $q_\perpp'=q_\perpp e^{\ell}$ and $q_z'=q_z e^{\int_0^\ell
\gamma(\ell')d\ell'}$,
where the function $\gamma\ofl$ is to be determined.
The dimension of \psif\ is just $(L^d)^{-1/2}$.  Note
that even when we rescale, \psif\ will rescale trivially because it
has no time dimensions.  \psif\ has no anomalous dimension
in our renormalization scheme, to leading order in $\epsilon=2-d$.

When doing the first momentum
shell integration, the coupling constants were
independent of length scale.  However, they then acquire a momentum
dependence because we have absorbed the large momentum
effects into them. The correct renormalized theory is a coupled set
of integral equations where the coupling constants are taken
to be scale dependent.  An alternative, but equivalent approach
is to integrate over a small momentum range where the coupling constants
are approximately fixed and then repeat the entire calculation iteratively.
This leads to the usual differential renormalization group
equations.

\subsec{Recursion Relations}

We now choose an infinitesimal momentum shell $e^{-\delta}$, and take the
limit $\delta\rightarrow 0$. This leads to
differential renormalization
group equations which can be integrated to produce the couplings appropriate
for a cutoff $q_\perpp<\Lambda e^{-\ell}$.  We use units such that $\Lambda=1$
in
the following.  The differential recursion relations are
\eqna\eREC{$$\eqalignno{
\der{D}&=D\left( -2+\gamma + {\lambda^2\over D}
{K_1(d-1)  + K_2\over K_1 K_2} {A_d\over 2d}\right)
&\eREC a\cr
\der{\lambda}&=\lambda\left( -1+\gamma\right)&\eREC b\cr
\der{v}&=v\left( -d+\gamma -{v\over 2D}A_d
+ {\lambda^2\over K_1 4D}A_d\right)
- {\lambda^4\over K_1^2 32D}A_d&\eREC c\cr
\der{K_1}&=K_1\left( d-2+\gamma\right)&\eREC d\cr
\der{K_2}&=K_2\left( d-2+\gamma\right)&\eREC e\cr
\der{K_3}&=K_3\left( d-\gamma\right)&\eREC f\cr
\der{\bar\mu}&=\bar\mu \gamma - \left({v\over D}
+ {\lambda^2\over 8K_1 D}\right)A_dD&\eREC g\cr}$$}
It is convenient to chose $\gamma\ofl$ so that the renormalized, rescaled
$D$ remains fixed at its initial value.
By examining \eREC{}\
we can see that the there are three dimensionless coupling constants
which come into this theory:
$\lb = \lambda (K_1(d-1)+K_2)^{1/2}(K_1K_2D)^{-1/2}$,
$\lan =\lambda (K_1D)^{-1/2}$ and $\vb = v/D$.

The quantity, \lb\ , controls the coupling
between the tangent field of the polymers and the nematic matrix.
The coupling, \lan\ , represents the interactions between density
fluctuations in the polymers and the the nematic.
Note that none of the dimensionless couplings depend on the
bend Frank constant, $K_3$.  For non-interacting polymers, $q_z\sim
q_\perpp^2$.
Thus we expect that the term ${1\over 2}K_3q_z^2\vert\dnb\vert^2\sim
q_\perpp^4$
and will be suppressed at very
long wavelengths.  This is why the bend elastic constant does not couple
to the theory at this order, though presumably it will at higher order.

The result that the geometric mean of the splay and twist Frank constants
comes into \lb\ is not unexpected.  In \Ra\ it
is shown that the effective rigidity of a polymer in a nematic
is the harmonic mean of its original rigidity and the induced
rigidity from the nematic.
We can understand the meaning of this harmonic mean by considering
the following static example of a single polymer.  If the polymer
deviates away from the $\hat z$ direction, the nematic can relieve
the stress in the {\sl same} time slice by a twist,
a splay or by a combination of the two.  In $d$ dimensions, there
are $d-1$ twist directions and only $1$ splay direction.  In the static
limit, where $q_z =0$, the nematic bends back to its
preferred direction over the transverse plane, whether
the relief is through a twist or a bend.
This allows us to estimate
the nematic energy cost in disturbing the polymer,
\eqn\enem{\delta F\propto \alpha^2K_1 + (d-1)({1-\alpha\over d-1} )^2K_2}
where $0\le\alpha\le 1$ measures the fraction of the
distortion relieved by a splay mode, and $(1-\alpha)/(d-1)$ is the
fraction of the distortion carried by each of the equivalent twist
modes.  Minimizing the energy with respect to $\alpha$, we
find that
\eqn\eNEM{\delta F\propto \alpha^2
\left({K_1K_2\over K_1(d-1) + K_2}\right),}
showing that the coupling to tangent fluctuations is given by \lb .

Finally, \lan\ is the coupling between the polymer and the splay
degrees of freedom of the nematic.  As we have seen, this coupling
comes about through density fluctuations of the polymers.
The effect of the nematic is to create an attraction between polymers,
not unlike the Van der Waals attraction between neutral bodies.

\subsec{Fixed Points and Flows}

Our theory is described by the full space of all the
coupling constants.  We can describe the flow of
the theory towards a stable theory by analyzing its fixed
point structure in this space.  The behavior of the Frank
constants is trivial, amounting to a mere rescaling. We assume, moreover,
that the chemical potential has been adjusted to the critical point
which describes the critical point of the theory \Ri .

We will analyze the flow in terms of the variables described above,
namely \vb\ , \lan\ and \lb\ near $d=2$. From \eREC{}\ we find:
\eqna\eDIMEN{$$\eqalignno{
\derr{\vb} &=\vb\bigg[\epsilon - {\vb\over 4\pi} +(\lan^2 -
\lb^2) {1\over 8\pi}\bigg] - {\lan^4\over 64\pi}&\eDIMEN a\cr
\derr{\lb} &=\lb\bigg({\epsilon\over 2} - {\lb^2\over 16\pi}\bigg)&\eDIMEN b\cr
\derr{\lan} &=\lan\bigg({\epsilon\over 2} - {\lb^2\over 16\pi}
\bigg)&\eDIMEN c\cr}$$}
Note that \lan\ is slaved to \lb\ in the sense that their ratio
is independent of $\ell$.
If we let $u=\vb - (1/4)\lan^2$, the recursion relation for $u\ofl$ depends
only on $\lb^2$.
\eqn\eU{
\derr{u} = u\left(\epsilon - {\lb^2\over 8\pi} - {u\over 4\pi}\right).}
so that it suffices to consider flows in the space of $u\ofl$ and $\lb^2\ofl$.
The subspace $\lb^2=0$ is the theory considered by Nelson and Seung \Ri\ for
flux lines, while the subspace $u=0$ is the non-interacting theory
considered by de Gennes \Ra .  By examining the flow to the fixed point, we
can decide which theory dominates in the long wavelength limit.

Since the dimension of interest is $d=2$, we discuss the flows for $\epsilon
=0$.  In this case we can solve \eDIMEN{b}\ for $\lb^2\ofl$,
\eqn\eflowoo{
\lb^2\ofl={\lb^2(0)\over \left[1+{\displaystyle{\lb^2(0)\over
8\pi}}\ell\right]}
\mathrel{\mathop\sim_{\ell\rightarrow\infty}}{8\pi\over\ell}
}
and then solve for $u$,
\eqn\eflowooo{
u\ofl= {u(0)\over \left(1+{\displaystyle{\lb^2(0)\over 8\pi}}\ell\right)\left[
1+{\displaystyle{2u(0)\over\lb^2(0)}}\ln\left(1+{\displaystyle{\lb^2\over
8\pi}}\ell\right)\right]}
\mathrel{\mathop\sim_{\ell\rightarrow\infty}}{4\pi\over\ell\ln\ell}}
if $\lb(0)\ne 0$.  If $\lb(0)=0$, $u\ofl$ is given asymptotically by
$4\pi/\ell$, in agreement with the results in \Ri .  The flows are illustrated
in Figure 13.

If $u_0>0$, the flows go the origin in the $(\lb^2,u)$-plane.  For nonzero
$\lb^2(0)$, they come into the origin tangent to the $\lb^2$ axis.  This
means that the logarithms that de Gennes discussed for a {\sl single}
polymer dominate the logarithms associated with interpolymer interactions.
If, on the other hand, $u_0<0$, the flows will still run to $\lb^2=0$, but
$u$ will run off to $-\infty$.  Since $u$ is the coefficient of
$\rho^2$ in the hydrodynamic theory, this flow will push the system
through a gas-liquid phase transition, since presumably there
are higher order terms in both the hydrodynamic and boson language,
corresponding to many-polymer interactions.

Finally, we can calculate the logarithmic corrections to the wandering
of a single polymer in the dilute limit,
\eqn\er{
\vev{\vert \rb(L)-\rb(0)\vert^2} =
{\int d\rb\,r^2\vev{\psi(\rb,L)\psi^*({\bf 0},0)}\over
\int d\rb\,\vev{\psi(\rb,L)\psi^*({\bf 0},0)}}.}
In order to do this, we observe that
\eqn\eRENORM{
\vev{\psi(\rb,z)\psi^*({\bf 0},0)}_{\ell=0} =
e^{d\ell}\vev{\psi(e^{-l}\rb,e^{-\int_0^\ell \gamma(\ell')\ell'}z)
\psi^*({\bf 0},0)}_{\ell} .}
We have chosen $\gamma\ofl$ so that $D$ will remain fixed at every scale, {\sl
i.e.},
$\gamma\ofl = 2-\lb^2\ofl /8\pi$.  This choice is arbitrary and,
of course, does not affect the expressions for
physical quantities.

We now choose $\ell^*$ such that
$z'=e^{-\int_0^{\ell^*}\gamma(\ell')d\ell'}z=a_0$,
where $a_0$ is the persistence length of a single, {\sl directed} polymer.
As $z\rightarrow\infty$, $\ell^*$ becomes large enough
so that we have flowed very close to the stable fixed point in Figure 13.
Since \lb , \lan\ and \vb\ are small near this fixed point, we can
simply use the zeroth order term in perturbation theory
to calculate the two-point function.  We have
\eqn\ezero{\eqalign{
\vev{\vert\rb(z)-\rb(0)\vert^2}_{\ell=0}&=
e^{2\ell^*}\vev{\vert\rb'(z')-\rb'(0)\vert^2}_{\ell=\ell^*}\cr
&=
e^{2\ell^*}
{
\int d^2\!r'\, (r')^2\vev{\psi(\rb',z')\psi^*({\bf 0},0)}_{\ell=\ell^*}\over
\int d^2\!r'\,\vev{\psi(\rb',z')\psi^*({\bf 0},0)}_{\ell=\ell^*}
}\cr
&= 2dDe^{2\ell^*}z'=2dDze^{\int_0^{\ell^*}[2- \gamma(\ell')]d\ell'}\cr}}
Substituting $\lb^2\ofl/8\pi$ for $2-\gamma\ofl$, we can integrate \eflowoo\
when $d=2$ and find
\eqn\ezerozero{
\vev{\vert\rb(z)-\rb(0)\vert^2}_{\ell=0}
=4Dz\left[1+{\displaystyle{\lb^2(0)\over 8\pi}}\ell^* \right]
}
Our choice of $\ell^*$ amounts to choosing
$\ell^*\approx\half\ln(z/a_0)$.  Substituting this into \ezerozero\
and writing \lb\ in terms of the original couplings (with $\lambda=1$, as
it was originally)
\eqn\ezerozerozero{
\vev{\vert\rb(z)-\rb(0)\vert^2}
=4Dz+{K_1+K_2\over 4\pi(K_1K_2)}z\ln\left({z\over a_0}\right)
}
again resulting in a logarithmic correction to wandering.  Comparison
with de Gennes' result \eIIvii\ shows that the logarithmic
correction to wandering is only half as large as would be predicted by
the simple argument in Section 2.

The correlation $\vev{\psi(\rb,z)\psi^*({\bf 0},0)}$
is the probability distribution for the wandering of a {\sl single}
polymer only in the dilute limit.  This correlation function represents
inserting a polymer at
$({\bf 0},0)$ and removing a polymer at $(\rb,z)$, but there is no
constraint that it be the same polymer.  However, if the system is
sufficiently dilute, the likelihood of polymers swapping their heads and tails
is small.  In the above derivation, we halted our renormalization
group iteration when $z\sim a_0$.  However, we can also stop
iterating when the polymer density, $\rho_0$, becomes of the order of the
$\Lambda^2$.
Since $\Lambda^{-1}$ is of order the monomer thickness, we can then
apply the hydrodynamic theory of Section 5.

For fixed $z$, we can always make the system dilute enough so that as
we follow the renormalization group trajectory $z(\ell)\sim a_0$
{\sl before} $\rho(\ell)\sim\Lambda^2$, where
$z\ofl=ze^{-\int_0^\ell\gamma(\ell')d\ell'}z$ and $\rho(\ell)=
e^{2\ell}\rho_0$.  In this regime the wandering
is given by \ezerozerozero .  However, as the density increases, we
come to a point where $\rho\ofl\sim\Lambda^2$ {\sl before} $z\ofl\sim a_0$.
Now
we must choose $\ell^*$ so that $\rho(\ell^*)\Lambda^2=1$.  In this case
\ezerozerozero\ will cross over to
\eqn\ezerozerozeroo{
\vev{\vert\rb(z)-\rb(0)\vert^2}
=4Dz+{K_1+K_2\over 4\pi(K_1K_2)}z\ln\left({1\over \rho_0\Lambda^2}\right)
}
This result will hold when ${z\over a_0}>{1\over \rho_0\Lambda^2}$.  In this
very long polymer regime, the interpolymer interactions destroy
the logarithmic correction to wandering.  Each time the polymers
wander into each other, the random walk is reset, and thus the logarithm
does not build up along their length.

\newsec{Effects of Hairpins}

In the preceding sections we considered polymers without
hairpin configurations.  As in Section 2.3,
we can account for hairpins by adding a term to the action \ethreeparts ,
namely
\eqn\epins{S\rightarrow S-{w\over 2}\int dzd^2\!r
[\psi^2+(\psi^*)^2]}
The coupling $w\propto\exp(-\epsilon_h/\kb T)$, where the hairpin energy
$\epsilon_h$ is related to the coupling constants in \eIii\ by
$\epsilon_h=\bo{\sqrt{g\kappa}}$ -- see Section 2.3.  Upon repeating the
analysis
of Section 3.2, we see that these terms create and destroy {\sl pairs}
of polymer lines and so add hairpins to the theory.

Upon carrying out perturbation theory in $v$ and $\lambda$,
we find that
unphysical diagrams appear in our theory.  By allowing hairpins, we now
also include loops of interacting polymers as in
Figure 14a.  However, following de Gennes
\ref\rdegnn{P.G. de Gennes, Phys. Lett. A~{\bf 38} (1972) 339.} ,
we can eliminate closed
loops of polymers by replicating $\psi(\rb,z)$ $M$ times, and then taking the
limit
$M\rightarrow 0$.  This could also have been done in our earlier analysis,
although it is unnecessary, due to the retarded nature of the propagators.
In the theory without hairpins but with free ends, one might think that
the free ends generate an effective hairpin term as in Figure 14b.  However,
the effective hairpin strength is $\bo{h^2}$, and makes good physical sense.
With free ends present, a long polymer may interact with two short
polymers, simulating the effect of an intervening closed loop polymer.  We must
be certain that taking $M\rightarrow 0$ preserves these graphs.
Consider the action for $M$ polymer fields coupled to a source (without
explicit hairpins),
\eqn\eREPL{
S=\sum_{\alpha=1}^M\int dzd^2\!r\,\left\{
\eqalign{
&\psi^*_\alpha\left(\partial_z-D\nablab_\perpp^2 -\bar\mu\right)\psi_\alpha
+{v\over 2}\sum_{\beta=1}^M \psi^*_\alpha\psi^*_\beta\psi_\alpha\psi_\beta\cr
&+ {\lambda\over 2}\dnb\cdot\left(\psi^*_\alpha\nablab_\perpp
\psi_\alpha -\psi_\alpha\nablab_\perpp\psi^*_\alpha\right)\cr
&- h_\alpha\left(\psi^*_\alpha+\psi_\alpha\right)\cr}\right\} +
{F_n[\dnb]\over\kb T}.}
By changing the replica basis, we can choose to only allow
the $\alpha=1$ direction to have a source term.  Because of the preferential
status of this direction, the ``effective'' closed loops will be \bo{1},
as opposed to \bo{M}\ in the case of real closed loops.  Thus
taking $M\rightarrow 0$ will not alter the results found earlier for free ends.

In the dense phase, hairpins lead to effects very similar to
those discussed for free ends in Section 4.  To see
how hairpins affect the transition near $\bar\mu=0$,
We first set $\psi\equiv(\psi_1+i\psi_2)
/\sqrt{2}$, and note that the action takes the form
\eqn\ediag{\eqalign{S= &{F_n[\dnb]\over\kb T}+\cr
&\int dzd^2\!r\,\left\{
\eqalign{
&i\psi_{\alpha\tiny 1}\partial_z
\psi_{\alpha\tiny 2}+ \half\psi_{\alpha\tiny
1}[-D\nablab_\perpp^2-(\bar\mu-w)]\psi_{\alpha\tiny 1}\cr&
+ \half\psi_{\alpha\tiny 2}[-D\nablab_\perpp^2-(\bar\mu+w)]\psi_{\alpha\tiny
2}+{v\over 8}(\psi_{\alpha\tiny 1}^2+\psi_{\alpha\tiny 2}^2)
(\psi_{\beta\tiny 1}^2+\psi_{\beta\tiny 2}^2)\cr
&+i{\lambda\over 2}\dnb\cdot(\psi_{\alpha\tiny
1}\nablab_\perpp\psi_{\alpha\tiny 2}
-\psi_{\alpha\tiny 2}\nablab_\perpp\psi_{\alpha\tiny 1})\cr}\right\}\cr}}
where we have followed the usual summation convention.
Recursion relations can be constructed as before for finite $M$, resulting in
\eqna\eRECN{$$\eqalignno{
\der{\bar\mu}&=\bar\mu \gamma - \left({\bar v\over 2\pi}(M+1)
+ {\lan^2\over 8\pi}\right)D&\eRECN a\cr
\der{w}&=w\left(\gamma - {u\over 4\pi}\right)&\eRECN b\cr}$$}
where $u=v-\lan^2/4$ was defined in the previous section. The
recursion relations for the other variables remain the same as in
\eREC{a-f} .
Upon taking $M\rightarrow 0$ we see that only the recursion relation for
$\bar\mu$ is different from \eREC{g}, and that (since $\gamma=2$) $w$ is a
strongly relevant perturbation.
Although $\bar\mu$ has a different recursion relation, this has
no effect on the universal long wavelength properties. The $M\rightarrow 0$
limit amounts to removing the second
graph in Figure 7c, while keeping the first graph.  As before,
the surviving graph can be absorbed into a change in the chemical potential.

Suppose our system is very dilute, {\sl i.e.} $\bar\mu\roughly{<}0$
and $w>0$, then $\psi_{\tiny 2}$ will condense.  We can
follow $\bar\mu+w$ via the recursion relations \eRECN{} .
The coupling $\bar\mu+w$ grows rapidly under iteration, until
it becomes large enough so that we can integrate out the massive
modes corresponding to the $\{\psi_{\alpha\tiny 2}\}$.  The
resulting theory involves only $\psi_{\alpha\tiny 1}$ and $\dnb$,
\eqn\eiso{S_{\rm eff}=
\int dzd^2\!r_\perpp\,\left\{\eqalign{
&{1\over 2}(\partial_z\psi_{\alpha\tiny 1})^2 + {D\over
2}(\nablab_\perpp\psi_{\alpha\tiny 1})^2
-{\bar\mu_1\over 2}\psi_{\alpha\tiny 1}^2\cr
&+v\psi_{\alpha\tiny 1}^2\psi_{\beta\tiny 1}^2
+ {\lambda^2\over 4}(\nablab_\perpp\psi_{\alpha\tiny 1}\cdot\dnb)^2\cr
&+ \lambda(\nablab_\perpp\psi_{\alpha\tiny 1}\cdot\dnb)
\partial_z\psi_{\alpha\tiny 1}\cr}\right\}+{F_n[\dnb]\over\kb T}.}
with $\bar\mu_{\tiny 1}=\bar\mu-w$.  We now have a theory in which directed
propagators are replaced by
merely anisotropic gradient couplings -- the meandering in the parallel
direction scales just as the meandering in the the perpendicular
direction with different proportionality constants.
The couplings shown in \eiso\ only serve as a caricature of those found
in the theory -- they will be changed by numerical
factors depending on at which point we
integrate out $\psi_{\tiny 2}$.  As we go through
the dilute-dense phase transition by letting $\bar\mu_{\tiny 1}$ change sign,
the theory describes a quasi-isotropic polymer melt with self avoidance in the
limit $M\rightarrow 0$ \refs{\rnewc,\rdegnn}.

The dilute limit of directed polymer melts {\sl without} hairpins
is described by an XY-like critical point (with a diffusive
propagator), as indicated by Figure
8.  The change in this phase transition induced by hairpins can be
summarized by considering the mean field diagram (see Figure 15) associated
with the polynomial part of \ediag .  We imagine holding $v$ and $w$ fixed, and
varying the chemical potential $\bar\mu$ in the plane of $r_1=-\bar\mu+w$ and
$r_2=-\bar\mu-w$.  When $w= 0$, the critical point, at $r_1=r_2=0$ is just
that considered in Figure 8.  When $w\ne 0$, however, the trajectory
as $\bar\mu$ varies passes through one of the {\sl Ising}-like critical
lines with a quasi-isotropic effective propagator (see \eiso ),
on the $r_1-$ and $r_2-$ axes.  In a similar fashion, the line describing
the XY-like symmetry of the dense phase for $r_1=r_2<0$ becomes
unstable to Ising-like dense phases when $w=0$.

The above discussion ignores the coupling to the nematic field.  This
theory is similar to the compressible Ising model \ref\rHalp{
D.J. Bergman and B.I. Halperin, Phys. Rev. B{\bf 13}, 2145 (1976).}
though the marginal operator present in that theory is not present
in ours.  This is guaranteed by the underlying nematic symmetry
of our theory.  Note that the original field theory is invariant under
$\dnb\rightarrow-\dnb$, $\psi\leftrightarrow\psi^*$ and $z\rightarrow -z$.
This prevents a term such as $(\nablab_\perpp\!\cdot\!\dnb)\psi_{\tiny 1}^2$,
the usual
coupling of the Ising model to an underlying elastic lattice.  The
couplings which do appear do not affect the above arguments, and we
are left with an $M$ component Ising model as $M\rightarrow 0$, reproducing
de Gennes' theory of isotropic polymers.  Note that the upper critical
dimension
of this Ising-like transition is $d_c=4$, as opposed to the result $d_c=2+1=3$
appropriate when $w=0$.

\newsec{Acknowledgments}

It is a pleasure to acknowledge helpful conversation with L. Balents,
T. Hwa, M. Goulian and R.B. Meyer during the course of this investigation.
We also benefited from the comments of an anonymous referee.
One of us (RDK), would like to acknowledge the support of a National Science
Foundation Graduate Fellowship.  This work was supported by the National
Science Foundation, through Grant DMR91-15491 and through the
Harvard Materials Research Laboratory.

\appendix{A}{Consequences of Rotational Invariance}

The full rotationally invariant coupling between the nematic and the
polymer is given by
\eqn\eAA{
F=g\int ds\,\left[ 1-\left({d\vec R(s)\over ds}\cdot \vec n\right)^2\right]
+F_n[\dnb],}
where $s$ is the arc-length along the polymer.  We now consider the following
transformations of our fields (leaving all other fields the same),
\eqn\eAAA{\eqalign{
n_x&\rightarrow n_x+v_x\cr
n_y&\rightarrow n_y+v_y\cr
R_z(s)&\rightarrow R_z(s) - v_xR_x(s) - v_yR_y(s)\cr},}
where ${\bf v}=(v_x,v_y)$ is a constant vector in the plane
perpendicular to $\hat z$.
It is easy to check that \eAA\ is invariant under these transformations if
$H$, the magnetic field, is $0$.
In terms of the small deviations defined in Section 2,
this transformation, to linear order in $\bf v$ and
the fields is
\eqn\eeieio{
\dnb'(\rb'(z'),z')=\dnb(\rb(z),z)+{\bf v}}
and
\eqn\aeiou
{{d\rb'(z')\over dz'}={d\rb(z)\over dz}}
and thus to linear order in ${\bf v}$
\eqn\echange{F'=F+g\int dz\,{\bf v}\cdot\left({d\rb(z)\over
dz}-\dnb(\rb(z),z)\right) }
However, we may absorb the shift in $\dnb$ into a redefinition of the field, so
as to eliminate the effect of the transformation and find that $F$ is invariant
under \eAAA .

Returning now to the coherent state field theory \eIIIix , we find that
it must be invariant under
\eqn\eopa{\eqalign{
\dnb'(\rb',z')&=\dnb(r,z)+{\bf v}\cr
\partial_z'&=\partial_z - {\bf v}\cdot\nablab_\perpp\cr}}
We now change the coefficient of $\psi^*\partial_z\psi$ from $1$ to some
constant value $\alpha$.
The transformations lead us to
\eqn\echangee{\eqalign{
S_\alpha'[&(\psi')^*,\psi',\dnb']\cr
&=S_\alpha[\psi^*,\psi,\dnb] + {\bf v}\cdot\int dzd^2\!r (-\alpha)
\psi^*\nablab_\perpp\psi + \half\left(\psi^*\nablab_\perpp\psi -
\psi\nablab_\perpp\psi^*\right)\cr
&=S_\alpha[\psi^*,\psi,\dnb] + (1-\alpha)\int dz d^2\!r {\bf
v}\cdot\psi^*\nablab_\perpp\psi  .\cr}}
Thus $S$ will only be independent of $\bf v$ if $\alpha=1$.

\appendix{B}{Derivation of Hydrodynamics}

In fact, the hydrodynamic theories described in Section 4 can be derived
from the more microscopic boson theory \ref\rhwa{
We acknowledge Terry Hwa for a similar derivation.}.
To this end, we begin with \eIIIxiii ,
taking $h=0$ to start,
\eqn\eBi{
S_p=\int dz d^2\!r\,\left[{D(\nablab_\perpp\rho)^2\over 4\rho} +
D\rho(\nablab_\perpp\theta)^2
+i\rho\partial_z\theta + i\rho\nablab_\perpp\theta\cdot\dnb -\bar\mu\rho
+{v\over 2}\rho^2\right]}
We now introduce a vector field $\P$, via a Hubbard-Stratonovich
transformation,
in order to eliminate the $D\rho(\nablab_\perpp\theta)^2$ term in \eBi .
We now have
\eqn\eBiia{Z_p=\int{\cal D}\P{\cal D}\rho{\cal D}\theta\,e^{-S'_p}}
with
\eqn\eBii{
S'_p=\int dz d^2\!r\,\left[\eqalign{
&{D(\nablab_\perpp\rho)^2\over 4\rho}
+i\rho\partial_z\theta + i\rho\nablab_\perpp\theta\cdot\dnb\cr& -\bar\mu\rho
+{v\over 2}\rho^2
+ D\rho \P^2 +2iD\rho\P\cdot\nablab_\perpp\theta\cr}\right]}
If we integrate out $\P$ we return to the original action $S_p$.  However, we
can now integrate over $\theta$.  Since it appears but linearly in $S'_p$,
its integration results in a delta-functional
\eqn\eBiii{
{\cal Z}=\int{\cal D}\P{\cal D}\rho
e^{-S_H}\delta[\partial_z\rho+\nablab_\perpp\cdot
(\rho\dnb+2D\P)]}
where
\eqn\eBiv{
S_H=\int dzd^2\!r\left[D\rho\P^2 +{D(\nablab_\perpp\rho)^2\over 4\rho}
-\bar\mu\rho
+{v\over 2}\rho^2\right].}
Thus we have traded interactions between $\theta$ and $\dnb$ for
a constraint relating $\rho$, $\P$ and $\dnb$.
We must now identify the physical meaning of $\P$.  In the case
$\dnb=0$, the constraint in \eBiii\ becomes
\eqn\eBv{
\partial_z\rho +\nablab_\perpp\cdot(2D\rho\P)=0}
which suggests that $2D\rho\P$ is the ``tangent'' field introduced
in Section 5.  However with the nematic field included, we would have
\eqn\eBvi{
\partial_z\rho+\nablab_\perpp\cdot(2D\rho\P+\rho\dnb)=0 .}
Returning to the path integral for a single polymer \eIIi , we
identify the Euclidean Lagrangian as
\eqn\eBvii{
L={1\over 4D}\left({d\rb\over dz}-\dnb\right)^2}
Leading to the Euclidean momentum
\eqn\eBviii{
{\bf p}=i{\partial L\over \partial\dot r}={i\over 2D}\left({d\rb \over
dz}-\dnb\right)}
so that
\eqn\eBix{
{d\rb\over dz}=2Di{\bf p} + \dnb .}
Apparently, $\P$ is the field associated with $i{\bf p}$.  Letting $\bf v$
be the field equivalent of the velocity, we have
\eqn\eBx{
{\bf v}=2D\P+\dnb ,}
so our constraint becomes
\eqn\eBxi{
\partial_z\rho+\nablab_\perpp\cdot(\rho{\bf v})=0}
Note that $\rho{\bf v}={\bf t}$ is the ``momentum'' field of our quantum
bosons, which
we have called the ``tangent'' field in the previous sections.  We now
replace $\P$ by $({\bf v}-\dnb)/2D$ in \eBiv .  Upon expanding
$\rho$ around some mean field value $\rho_0$, we are lead to the
hydrodynamics of Section 5.

If we were to add sources to the theory, \eBi\ is changed by
\eqn\eBxii{
\delta S_p=\int dz d^2\!r\, 2h\sqrt{\rho}\cos\theta}
We may expand this term in powers of $\theta$ if the fluctuations
in $\theta$ are small, which they will be if the polymers are
sufficiently long.  To lowest order \eBxii\ is
\eqn\eBxiii{\delta S_p=\int dzd^2\!r\,h\sqrt{\rho}\theta^2}
Now upon integrating out $\theta$, we no longer have a $\delta$ functional
as in \eBiii , but instead
\eqn\eBxiv{
Z=\int{\cal D}{\bf v}{\cal D}\rho\, e^{-S_H}}
with
\eqn\eBxv{
S_H=\int dzd^2\!r \left\{\eqalign{
&D\rho({\bf v}-\dnb)^2 + {D(\nablab_\perpp\rho)^2\over 4\rho}
-\bar\mu\rho +{v\over 2}\rho^2 \cr&+ {1\over
4h\sqrt{\rho}}\left[\partial_z\rho+\nablab\cdot
(\rho{\bf v})\right]^2\cr}
\right\}}
resulting in the hydrodynamic theory of finite length polymers \efivefive .
There
we had the term
\eqn\eoldstuff{\half {G\over \kb T}(\partial_z\rho+\nablab\cdot{\bf t})^2.}
Recalling $G=\kb T\ell/2\rho_0$ and $\ell=\rho_0^{-1/2}/h$, we have
$\half(G/\kb T)=(4\sqrt{\rho_0}h)^{-1}$, as required.

The integration over $\theta$ in \eBii\ needs a more careful analysis.  Indeed,
it will constrain the number of lines passing through a surface to be integer.
When changing variables from $\psi$ and $\psi^*$ to $\rho$ and $\theta$, theta
is only determined modulo $2\pi$.  Because of this, we can consider
field configurations in $\theta$ which increase by $2\pi$ as we go around a
particular line.  This vortex line is analogous to a point vortex in the
two-dimensional $XY$ model.  We can now consider a closed vortex loop (the
three dimensional analog of a vortex-antivortex pair) where $\theta$ jumps
from $0$ to $2\pi n$ as we traverse a surface with the vortex loop as the
boundary (a so-called ``branch'' disc).  Rewriting the $\theta$ dependent part
of \eBii , we have, for a volume $\Sigma$ bounded by the surface
$\partial\Sigma$,
\eqn\eBBB{
i\int_\Sigma dV\,\left(\rho\partial_z\theta+{\bf t}\cdot
\nablab_\perpp\theta\right)=i\int_{\partial\Sigma}d\vec S\cdot\vec T\theta
-i\int_\Sigma dV\,\theta\left(\partial_z\rho + \nablab_\perpp\cdot{\bf
t}\right)}
where $\vec T=(t_x,t_y,\rho)$ is the three-dimensional ``particle'' current.
The volume integral on the right hand side of \eBBB\ will lead to
the constraint \eBxi\ in the volume $V$, when $\theta$ is functionally
integrated over.  The surface integral contains the new constraint.

Consider a vortex line which lies on the surface $\partial\Sigma$ (see Figure
16).  If we now split $\partial\Sigma$ into the parts above and below the
surface
of discontinuity, $\partial\Sigma_u$ and $\partial\Sigma_d$, respectively,
the surface integral is
\eqn\eBBBx{
2\pi in\int_{\partial\Sigma_u}d\vec S\cdot\vec T + i\int_{\partial\Sigma}
d\vec S\cdot\vec T\theta_s}
where $\theta_s$ is the smoothly varying part of $\theta$.  In the same way
that we generated the equation of continuity, the second integral will
constrain the total flux through the closed surface $\partial\Sigma$ to
be $0$ (no free ends.).  However, upon summing over all possible values of
$n$, the first integral in \eBBBx\ will constrain the ``flux'' of lines
passing through the surface $\partial\Sigma_u$ to be an {\sl integer}.
If polymers are dense and entangled, this set of nonlocal constraints
should be unimportant in the hydrodynamic limit, where the fluctuations
in $\theta$ are small.
\vfill\eject
\vbox{\offinterlineskip
\hrule
\halign{&\vrule#&\strut\quad#\quad\cr
height2pt&\omit&&\omit&&\omit&&\omit&\cr
&\omit&&Mechanism For\hfil&&Free Ends\hfil&&Hairpins\hfil&\cr
&System\hfil&&Alignment\hfil&&Important?\hfil&&Important?&\cr
height2pt&\omit&&\omit&&\omit&&\omit&\cr
\noalign{\hrule}
height2pt&\omit&&\omit&&\omit&&\omit&\cr
&Polymer\hfil&&Spontaneously\hfil&&\hfil Yes\hfil&&\hfil Yes\hfil&\cr
&Nematics\hfil&&Broken Symmetry\hfil&&\omit&&\omit&\cr
height2pt&\omit&&\omit&&\omit&&\omit&\cr
\noalign{\hrule}
height2pt&\omit&&\omit&&\omit&&\omit&\cr
&Polymers In A\hfil&&Induced By\hfil&&\hfil Yes\hfil&&\hfil Yes\hfil&\cr
&Nematic Solvent\hfil&&Solvent\hfil&&\omit&&\omit&\cr
height2pt&\omit&&\omit&&\omit&&\omit&\cr
\noalign{\hrule}
height2pt&\omit&&\omit&&\omit&&\omit&\cr
&Ferro- and\hfil&&External\hfil&&\hfil Yes\hfil&&\hfil No\hfil&\cr
&Electrorheological\hfil&&Fields\hfil&&\omit&&\omit&\cr
&Fluids\hfil&&\omit&&\omit&&\omit&\cr
height2pt&\omit&&\omit&&\omit&&\omit&\cr
\noalign{\hrule}
height2pt&\omit&&\omit&&\omit&&\omit&\cr
&Flux Lines In\hfil&&External\hfil&&\hfil No\hfil&&\hfil No\hfil&\cr
&High Temperature\hfil&&Fields\hfil&&\omit&&\omit&\cr
&Superconductors\hfil&&\omit&&\omit&&\omit&\cr
height2pt&\omit&&\omit&&\omit&&\omit&\cr
\noalign{\hrule}
height.36truein width0pt&\omit&width0pt&\omit&width0pt&\omit&width0pt&\omit
&width0pt\cr
\multispan9 Table I.\qquad $\,$Characteristics of the
different physical systems\hfil\cr
height12pt width0pt&\omit&width0pt&\omit&width0pt&\omit&width0pt&\omit
&width0pt\cr
\multispan9 ~~~~~~~~~~~\qquad considered in this this paper.\hfil\cr
}}

\listrefs
\listfigs
\bye